\documentclass[twocolumn]{aastex701}

\usepackage{xspace}
\usepackage[version=4]{mhchem} 
\usepackage{hyperref} 
\usepackage{diagbox}

\newcommand{\eureka}{\texttt{Eureka!}\xspace}
\newcommand{\poseidon}{\texttt{POSEIDON}\xspace}
\newcommand{\pRT}{\texttt{petitRADTRANS}\xspace}
\newcommand{\picaso}{\texttt{PICASO}\xspace}
\newcommand{\firefly}{\texttt{FIREFLy}\xspace}

\received{\today}

\shorttitle{KELT-7\,b NIRISS}

\shortauthors{Schmidt et al.}

\begin{document}

\title{Mitigating Charge Migration in JWST NIRISS Reveals That KELT-7\,b is a Metal-enriched Ultra-hot Jupiter Orbiting a Young Metal-rich Star}

\author[0000-0001-8510-7365]{Stephen P.\ Schmidt}
\altaffiliation{NSF Graduate Research Fellow}
\affiliation{William H. Miller III Department of Physics and Astronomy, Johns Hopkins University, Baltimore, MD 21218, USA}
\email{sschmi42@jhu.edu}

\author[0000-0002-2739-1465]{Erin M.\ May}
\affiliation{Johns Hopkins Applied Physics Laboratory, Laurel, MD 20723, USA}
\email{erin.may@jhuapl.edu}

\author[0000-0003-3667-8633]{Joshua D.\ Lothringer}
\affiliation{Space Telescope Science Institute, 3700 San Martin Drive, Baltimore, MD 21218, USA}
\email{jlothringer@stsci.edu}

\author[0000-0003-0473-6931]{Patrick McCreery}
\affiliation{William H. Miller III Department of Physics and Astronomy, Johns Hopkins University, Baltimore, MD 21218, USA}
\email{pmccree2@jhu.edu}

\author[0000-0002-1624-3360]{Mei Ting Mak}
\altaffiliation{Croucher Postdoctoral Fellow}
\affiliation{Atmospheric, Oceanic, and Planetary Physics Department, University of Oxford, OX1 3PU, UK}
\affiliation{Department of Physics and Astronomy, Faculty of Environment, Science and Economy, University of Exeter, Exeter EX4 4QL, UK}
\email{martha.mak@physics.ox.ac.uk}

\author[0000-0001-7361-0828]{Myles Pope}
\affiliation{William H. Miller III Department of Physics and Astronomy, Johns Hopkins University, Baltimore, MD 21218, USA}
\email{mpope7@jhu.edu}

\author[0009-0003-1097-5792]{Harry Baskett}
\affiliation{Department of Physics and Astronomy, Faculty of Environment, Science and Economy, University of Exeter, Exeter EX4 4QL, UK}
\email{hb595@exeter.ac.uk}

\author[0000-0003-1622-1302]{Sagnick Mukherjee}
\altaffiliation{51 Pegasi\,b Postdoctoral Fellow}
\affiliation{School of Earth and Space Exploration, Arizona State University, Tempe, AZ 85287, USA}
\email{smukhe50@asu.edu}

\author[0000-0001-6050-7645]{David K.\ Sing}
\affiliation{William H. Miller III Department of Physics and Astronomy, Johns Hopkins University, Baltimore, MD 21218, USA}
\affiliation{Morton K. Blaustein Department of Earth \& Planetary Sciences, Johns Hopkins University, Baltimore, MD 21218, USA}
\email{dsing@jhu.edu}

\author[0000-0002-9030-0132]{Katherine A.\ Bennett}
\affiliation{Morton K. Blaustein Department of Earth \& Planetary Sciences, Johns Hopkins University, Baltimore, MD 21218, USA}
\email{kbenne50@jhu.edu}

\author[0000-0002-4701-8916]{Arika Egan}
\affiliation{Johns Hopkins Applied Physics Laboratory, Laurel, MD 20723, USA}
\email{arika.egan@jhuapl.edu}

\author[0000-0002-3263-2251]{Guangwei Fu}
\affiliation{William H. Miller III Department of Physics and Astronomy, Johns Hopkins University, Baltimore, MD 21218, USA}
\email{gfu@jhu.edu}

\author[0000-0002-5113-8558]{Daniel P.\ Thorngren}
\affiliation{William H.\ Miller III Department of Physics \& Astronomy,
Johns Hopkins University, 3400 N Charles Street, Baltimore, MD 21218, USA}
\email{dpthorngren@jhu.edu}

\author[0000-0002-4997-0847]{Duncan A. Christie}
\affiliation{Max Planck Institute for Astronomy, Königstuhl 17, D-69117 Heidelberg, Germany}
\affiliation{Department of Physics and Astronomy, Faculty of Environment, Science and Economy, University of Exeter, Exeter EX4 4QL, UK}
\email{christie@mpia.de}

\author[0000-0001-5097-9251]{Carlos Gascón}
\affiliation{Space Telescope Science Institute, 3700 San Martin Drive, Baltimore, MD 21218, USA}
\email{cgascon@stsci.edu}

\author[0000-0002-6379-3816]{Le-Chris Wang}
\affiliation{Department of Astrophysical Sciences, Princeton University, 4 Ivy Lane, Princeton, NJ 08544, USA}
\email{lechris.wang@princeton.edu}

\author[0000-0002-1056-3144]{Lakeisha M. Ramos Rosado}
\affiliation{William H. Miller III Department of Physics and Astronomy, Johns Hopkins University, Baltimore, MD 21218, USA}
\email{lramosr1@jhu.edu}

\author[0000-0001-6707-4563]{Nathan J. Mayne}
\affiliation{Department of Physics and Astronomy, Faculty of Environment, Science and Economy, University of Exeter, Exeter EX4 4QL, UK}
\email{N.J.Mayne@exeter.ac.uk}

\author[0000-0002-0832-710X]{Natalie H. Allen}
\altaffiliation{NSF Graduate Research Fellow}
\affiliation{William H. Miller III Department of Physics and Astronomy, Johns Hopkins University, Baltimore, MD 21218, USA}
\email{nallen19@jhu.edu}

\author[0000-0003-4408-0463]{Zafar Rustamkulov}
\affiliation{IPAC, California Institute of Technology, MC 100-22, 1200 East California Boulevard, Pasadena, CA 91125, USA}
 \email{zafar@ipac.caltech.edu}

\author[0000-0003-3204-8183]{Mercedes López-Morales}
\affiliation{Space Telescope Science Institute, 3700 San Martin Drive, Baltimore, MD 21218, USA}
\email{mlopez-morales@stsci.edu}

\author[0000-0001-5761-6779]{Kevin C.\ Schlaufman}
\affiliation{William H. Miller III Department of Physics and Astronomy, Johns Hopkins University, Baltimore, MD 21218, USA}
\email{kschlaufman@jhu.edu}

\correspondingauthor{Stephen Schmidt}
\email{sschmi42@jh.edu}

\begin{abstract}
\noindent 
We present the first panchromatic JWST transmission spectrum of an ultra-hot Jupiter, combining NIRISS and NIRSpec observations to constrain KELT-7\,b's atmospheric properties. We show evidence for charge migration in our NIRISS observation between 1--1.5 $\mu$m, a wavelength range crucial to test for enhanced H$^-$ previously inferred from HST WFC3/IR G141 observations. We mitigate charge migration by fitting the ramp after extracting 1D stellar spectra at the group level. This ``late-ramp-fit'' method calculates KELT-7\,b's transmission spectrum between 1--1.5 $\mu$m to more closely match data less affected by the brighter-fatter effect at higher signal-to-noise. Using the transit-derived stellar mean density during stellar property inference reveals that KELT-7 is a $640\pm100$ Myr-old, $[\text{Fe}/\text{H}]=0.46\pm0.02$ star. Combined with NIRSpec and re-reduced WFC3/UVIS G280 data, our free retrieval analysis shows strong evidence for H$_2$O, CO$_2$, and TiO among high-temperature species, but not H$^-$ or clouds. Unaccounted-for systematics may therefore bias longer-wavelength WFC3/IR G141 transit depths shallower. Our free retrieval, two equilibrium retrievals, and self-consistent grid fit all prefer a high metallicity but find discrepant C/O ratios. Agglomerated together, we constrain a super-stellar $\text{M/H}=92^{+24}_{-23}\times$ Solar and C/O $\leq0.9$, suggesting enhanced metal accretion in the later stages of KELT-7\,b's formation. Our GCMs explain the observed lack of limb asymmetry with similar thermal structures between the two limbs, as well as superrotating jet-driven efficient horizontal mixing. The stark contrast between our panchromatic analysis and prior analyses on subsets of these data demonstrates the value of broad wavelength coverage for the comprehensive study of exoplanet atmospheres.

\end{abstract}


\submitjournal{AAS Journals}

\section{Introduction}
Ultra-hot Jupiters (UHJs), giant exoplanets with equilibrium temperatures $T_{\text{eq}} \gtrsim 2000$ K, offer some of the clearest views of exoplanet atmospheric physics.
At these temperatures cloud formation is limited to only a few species,
making their hot day sides free of clouds \citep{Parmentier18, Kitzmann18}.
Meanwhile, the vaporization of refractory metals into metal oxides and hydrides enables a more complete view of the chemical makeup of giant exoplanets \citep{Lothringer21, Lothringer25} relative to the cloudier views of cooler hot Jupiters \citep{Sing16}.
Their closeness to their host stars also makes UHJ systems some of the best laboratories for the study of star--planet interactions, like tidal decay \citep[e.g.,][]{Yee2020}, mass loss \citep[e.g.,][]{Haswell12}, anomalous heating \citep[e.g.,][]{Thorngren18, Schmidt26b}, and magnetic interactions \citep[e.g.,][]{Lanza09, Lanza11}.
This makes UHJs the most valuable class of planets for assessing the connections between atmospheric composition, atmospheric dynamics, and planet formation.

The James Webb Space Telescope (JWST) has revolutionized our view of UHJs through its unmatched precision and broad wavelength coverage, especially in conjunction with Hubble Space Telescope (HST) ultraviolet-visible light observations.
Emission spectroscopy of UHJs with JWST has shown additional evidence for thermal inversions \citep[e.g.,][]{Coulombe23, Pelletier26} that were predicted for highly irradiated planets \citep[e.g.,][]{Hubeny03, Fortney08} and previously discovered using HST eclipse data \citep{Evans17}.
Though TiO and VO, expected to be the main cause of thermal inversions, has been observed in UHJs through HST observations and high-resolution ground-based spectroscopy \citep[e.g.,][]{Sedaghati17, Edwards20, Fu21, Prinoth22, Pelletier23}, their presence has not yet been confirmed with JWST.
The presence of other refractory species like SiO has been confirmed in UHJ atmospheres with JWST though, enabling calculation of elemental ratios to assess formation scenarios \citep[e.g.,][]{EvansSoma25, Gapp25, Lothringer25}.
3D atmospheric analyses of UHJs with JWST have also found evidence for weaker longitudinal temperature gradients \citep{Challener25} and inefficient heat redistribution \citep{Splinter25}, in contrast with cloud cycling seen in cooler hot Jupiters \citep[e.g.,][]{Fu25, Mukherjee25}.
Panchromatic case studies of UHJs have now been performed on secondary eclipse spectra \citep[e.g.,][]{Kahle2026}; however, to date no UHJ transmission spectrum study has incorporated more than 1 JWST instrument into its analysis, limiting the view of these planets' atmospheres.

KELT-7\,b has recently joined the ensemble of UHJs observed by JWST and HST; however, discrepant findings have placed it strongly in tension with our understanding of typical UHJ atmospheres.
The first published transmission spectroscopy of KELT-7\,b was performed with HST's WFC3 using its IR/G141 filter.
With this data, \citet{Pluriel20} inferred evidence for H$_2$O and a high abundance of H$^-$ in the planet's atmosphere.
Further HST observations of KELT-7\,b at the shorter wavelengths offered by WFC3's UVIS/G280 filter suggested the presence of inhomogeneities on the surface of KELT-7 and confirmed the presence of H$^-$ seven orders of magnitude higher than predictions from equilibrium chemistry \citep{Gascon25}.
The first JWST observations of KELT-7\,b, providing near- to mid-infrared wavelength coverage via its Near Infrared Spectrograph \citep[NIRSpec;][]{NIRSpec1, NIRSpec2}'s G395H mode, again showed weak features that were interpreted as evidence for either a low-metallicity atmosphere or a high cloud deck muting molecular features \citep{Ahrer25}.
The JWST NIRSpec/G395H results were difficult to reconcile with the amount of H$^-$ inferred from the HST WFC3 IR/G141 data, resulting in an inconclusive story for the atmosphere of the planet.

With its 0.85-2.85 $\mu\text{m}$ wavelength coverage, JWST's Near Infrared Imager and Slitless Spectrograph \citep[NIRISS;][]{NIRISS1, NIRISS2} instrument bridges the gaps between these prior observations to offer a complete view of the molecular features in KELT-7\,b's ultraviolet to mid-infrared transmission spectrum.
However, due to KELT-7's bright apparent magnitude ($m_J = 7.739$), the KELT-7 system is close to the NIRISS Substrip 96 detector's brightness limit, resulting in additional systematic noise and broadening of the NIRISS point-spread function (PSF) due to accumulated charge.
This effect, also known as the ``brighter-fatter effect (BFE),'' has been observed to occur in near-infrared Hawaii-XRG (HXRG) HgCdTe photodiode arrays used by JWST's NIRISS, NIRSpec, and Near-infrared Camera (NIRCam) instruments \citep{Plazas18, Hirata20}, as well as in mid-infrared detectors like JWST's Mid-infrared Instrument (MIRI) \citep{Argyriou23}.
The net result is migration of charges from their observed locations, introducing nonlinearity that produces shallower transits in the brightest part of the detector.
Maximizing the science output of these observations therefore requires this charge migration to be addressed during the reduction process.

In this article we present a method to mitigate charge migration and unlock the full NIRISS transmission spectrum of KELT-7\,b.
We discuss this method in Section \ref{sec:chargemigration} and describe the standard steps of our reduction process in Section \ref{sec:reduction}; additionally, we reanalyze the HST WFC3 UVIS/G280 observations in Section \ref{sec:hst}.
To provide accurate inputs for our interpretation of the atmosphere of KELT-7\,b, we perform an updated characterization of the host star KELT-7 using archival observations in Section \ref{sec:stellar}.
We then describe our retrieval analysis of the panchromatic transmission spectrum in Section \ref{sec:retrievals} and perform 3D climate modeling of KELT-7\,b's atmosphere in Section \ref{sec:gcm}.
We discuss our results in Section \ref{sec:disc} and summarize our findings in Section \ref{sec:conclusion}.

\section{Mitigation of Charge Migration}\label{sec:chargemigration}
\begin{figure*}[ht!]
    \centering
    \includegraphics[width=\linewidth]{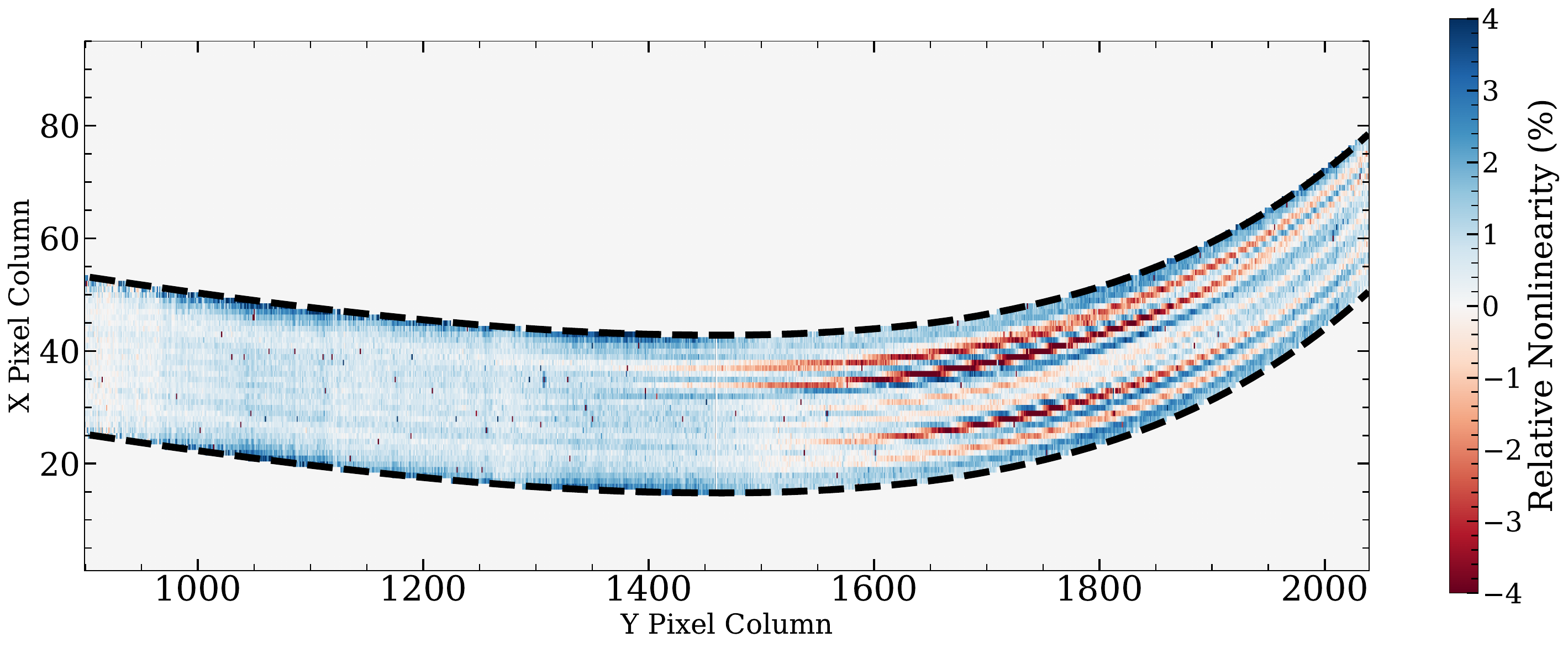}
    \caption{Temporal average of the relative nonlinearity in our KELT-7\,b NIRISS/SOSS observations after following the \texttt{jwst} pipeline up to and including the linearity step.
    Here we define the relative nonlinearity as the ratio between the differences between adjacent frames relative to 1, i.e., $(f_2 - f_1)/(f_1-f_0) - 1$ where 0 implies no nonlinearity.
    The color bar is cut off at $\pm4\%$ to highlight the effect of charge migration.
    We also apply a PSF mask to the data, showing the cutoffs as dashed black lines.
    There are several traces within the wider NIRISS/SOSS trace where there is significant nonlinearity in the negative direction.
    Adjacent to these traces on both sides are other traces with significant nonlinearity in the positive direction.
    This is an effect we would expect to see if vertical (i.e. spacial) charge migration is affecting our data, meaning that correction during the reduction process is necessary to accurately measure the transit depth in the brightest region of the detector.
    }
    \label{fig:chargemigration_avg}
\end{figure*}
The NIRISS instrument's full well capacity is roughly 105,000 e$^-$, i.e., $2^{16}$ Analog-to-Digital Units (ADU) at a gain of 1.61 e$^-$/ADU. However, several effects limit the ability of the instrument to accurately measure counts beyond certain thresholds below this limit.
Some of its pixels experience A/D saturation, which can decrease the theoretical maximum for the affected pixels by up to 20,000 e$^-$.
Nonlinearity effects also limit the maximum counts for a given pixel to be as low as 96,600 e$^-$, and pedestal/bias subtraction further decreases the upper limit to 68,000-92,000 e$^-$.
The final limiting effect occurs when adjacent pixels report highly differing counts, where charges in the high-count pixel alter the detector's electric field and cause additional photoelectrons to leak into neighboring pixels with lower charge.
This effect is known as charge migration and is hypothesized to be responsible for the ``brighter--fatter effect'' that alters the PSF in near-infrared detectors like NIRISS \citep{Hirata20}. For the NIRISS detector, charge migration has been observed from count rates starting at $\sim$23,300 e$^-$, about 1/2 well-depth, though the actual limit has been seen to depend on the filter.
While it sets a hard limit for NIRISS's aperture masking interferometry mode due to the nature of those observations, charge migration is not prohibitive for SOSS observations of brighter targets, though it causes nonlinearity in the resulting data when handled naively by the \texttt{jwst} pipeline\footnote{More information about the NIRISS detector's performance is available at \url{https://jwst-docs.stsci.edu/jwst-near-infrared-imager-and-slitless-spectrograph/niriss-instrumentation/niriss-detector-overview/niriss-detector-performance\#gsc.tab=0}.}.

When we initially investigated the observations, we noticed a nonlinearity in the wavelength range $1~\mu\text{m}\lesssim\lambda\lesssim1.5~\mu\text{m}$.
In the brightest regions of the detector ($>23,000$ e$^-$) we found clear evidence of the BFE, which we attribute to charge migration, as charges are seen to move from the brightest pixels to the nearby fainter pixels in the cross-dispersion direction. We illustrate this in Figure \ref{fig:chargemigration_avg}, where the count rate is seen to decrease in the brightest pixels and increase in the adjacent pixels above and below.
Based on this evidence, we hypothesized that the additional nonlinearity in the ramp fit we observed when following the standard steps of the \texttt{jwst} pipeline could be caused by vertical (i.e. spacial---see Figure \ref{fig:chargemigration_avg}) charge migration.

If a pixel is unable to record all of the photons it receives in the last group due to charge migration, then the photon measurements for that group in the brightest parts of the detector are lower than they are in reality.
This results in a lower measured baseline flux of the star at these wavelengths.
However, during a transit, the overall count rate is lower and therefore less affected by charge migration, impacting the out-of-transit flux more than the in-transit flux in a way that cannot simply be divided out.
When the planet occults the star, the measured dip in brightness is relatively less than it would be had the entirety of the star's baseline flux been captured, lowering the recorded transit depth at the affected wavelengths.
With only three groups per integration, incorrect photon measurements for even a single group will significantly lower the measured transit depth in the brightest regions of the detector.
It is therefore imperative to resolve the charge migration issue to successfully probe the full panchromatic transmission spectrum of KELT-7\,b.

If charge migration is indeed the cause of the nonlinearity, then collapsing the vertical axis, i.e. extracting the 1D stellar spectrum by summing the pixels within the trace, \emph{prior} to ramp fitting would allow the migrated charges to be counted as if they were in their ``correct'' locations.
This would result in an accurate transit depth that is in agreement with the groups unaffected by charge migration.
We therefore perform a modified version of the standard \texttt{jwst} data reduction pipeline that fits the ramp after summing each column of the detector to test this hypothesis.
Here we assume that the extracted pixels' nonlinearities are similar enough that extracting on the read pairs gives a sufficient approximation of a BFE correction.
Combining pixels at the group level to mitigate charge migration has been performed previously on MIRI data \citep{Deming2024}; here, we apply a similar methodology to NIRISS observations.
We call this modification the ``late-ramp-fit'' method, and execute the vertical charge migration test within the framework of two well-tested, open-source JWST spectrophotometric time series reduction pipelines: \firefly \citep{Rus22, Rus23, Sing24a} and \eureka \citep{Eureka!}.

Upon reaching and completing the linearity step, both of our reductions diverge from the standard \texttt{jwst} pipeline reduction steps.
We immediately perform the background subtraction, $1/f$ noise subtraction, and image cleaning steps for each individual group as if it were the finished output from Stage 1.
With these group-level extracted 1D spectra, we calculate \textit{read pairs}, or group differences\footnote{In other observational data reduction contexts, ``reads'' typically refer to samples up the ramp or by frames rather than group differences.}, and perform a 1$^{\text{st}}$-order \texttt{numpy} polyfit polynomial ramp fit with read pairs 1 and 2 (i.e., not including group 0 alone) and using the errors on each as weights, measuring the count rate in each pixel of the 1D spectrum in e$^-$/s.
The polynomial fit ignores any correlation that may arise from sharing group 1, as well as any correlations between the groups due to photon noise; however, we expect that this effect is diminished when averaging over many pixels.
This late-ramp-fit results in a single spectrophotometric time series that has corrected for spacial charge migration; at this stage, the data is in the same format and units as a traditional analysis.

\begin{figure*}
    \centering
    \includegraphics[width=\linewidth]{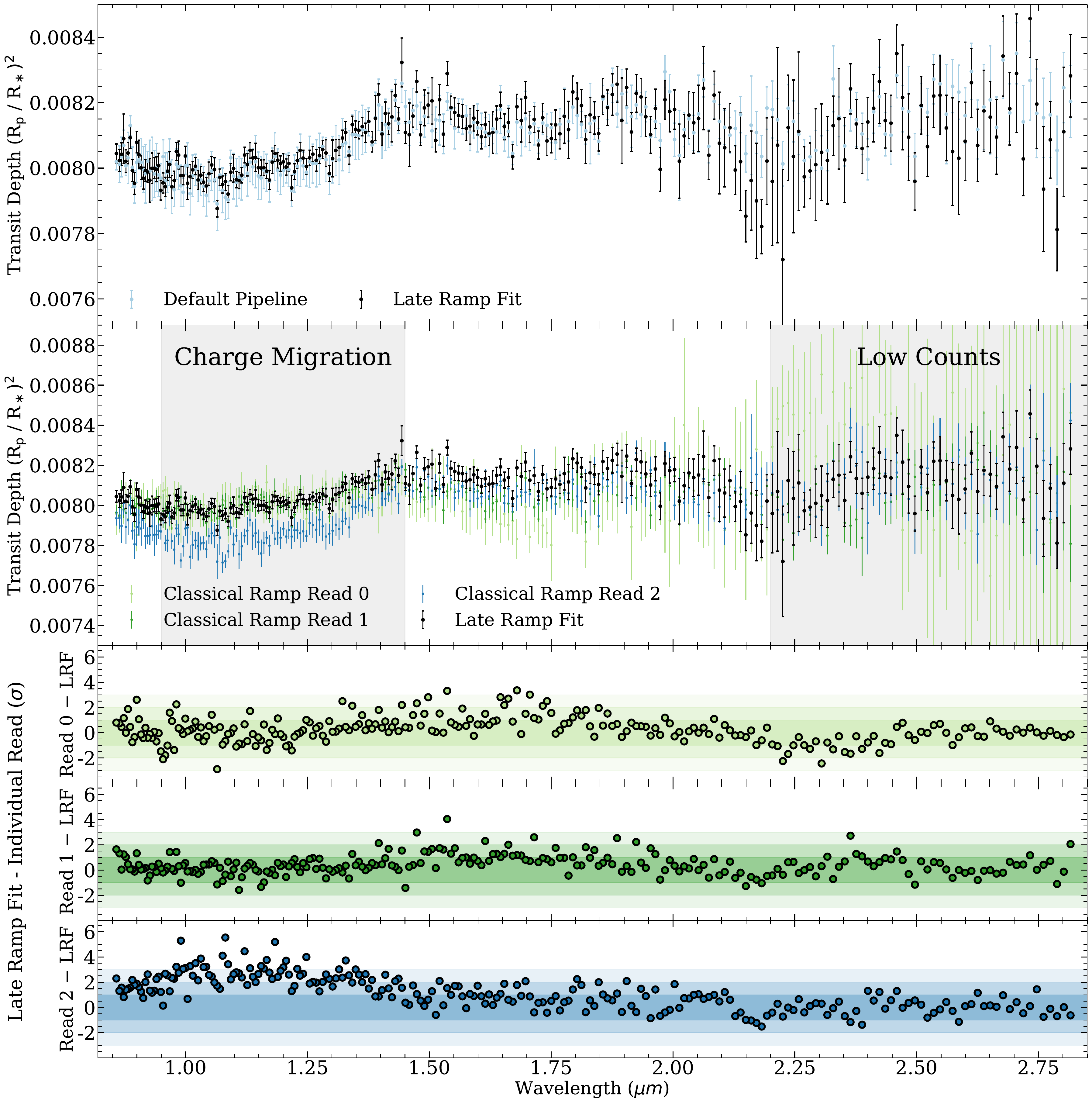}
    \caption{Comparison between \firefly reductions of the individual read pairs of our KELT-7\,b data and the resulting spectrum from our late-ramp-fit method. 
    We plot in the top panel as light blue error bars the transmission spectrum using the default \texttt{jwst} pipeline.
    In addition to the larger error bars, there is a very small downward bias in transit depth when not accounting for the BFE.
    We plot in the second panel as colored error bars the default-pipeline transmission spectra for Read pair 0 (light green; group 0), Read pair 1 (dark green; groups 0 \& 1), and Read pair 2 (dark blue; groups 1 \& 2), as well as our late-ramp-fit spectrum (larger black markers with error bar caps; all three groups). 
    We also highlight as the gray regions the charge migration-affected area (left) and the low-count area (right).
    We plot in the lower three panels the discrepancy between our late-ramp-fit spectrum and each read pair spectrum, with the marker face colors corresponding to the read pair in the upper panel.
    We show the 1-$\sigma$, 2-$\sigma$, and 3-$\sigma$ ranges on these panels as colored bands of increasing transparency centered at zero, with the color corresponding to the read pair.
    Read pairs 0 and 1 exhibit deeper transit depths between 1--1.5 $\mu$m with large error bars at the longest wavelengths.
    We note that at certain wavelengths the late-ramp-fit spectrum is discrepant from the individual read pair spectra.
    This is most prevalent with Read pair 0, which we attribute to flux-dependent reset bias.
    On the other hand, Read pair 2 (which is most affected by charge migration) has a shallower transit depth between 1--1.5 $\mu$m often discrepant by more than 3$\sigma$, but smaller error bars at longer wavelengths.
    Our method of mitigating charge migration allows our resulting spectrum to maintain higher signal-to-noise than the individual read pair spectra at all wavelengths without compromising the unaffected groups' transit depths at shorter wavelengths.}
    \label{fig:readcomparison}
\end{figure*}

To illustrate the correction and precision offered by the late-ramp-fit method, we show the \firefly spectra for each read pair in comparison to the final transmission spectrum in Figure \ref{fig:readcomparison}.
Between 1--1.5 $\mu$m, the first two read pairs yield consistent transit depths less affected by the BFE whereas the final read pair is corrupted by nonlinearity.
Conversely, at longer wavelengths the first two read pairs suffer from higher noise and large error bars, requiring all three groups in this wavelength range to maximize precision. 
Our late-ramp-fit method obviates this trade-off by producing a final spectrum that successfully captures both the accuracy of the unaffected early read pairs at short wavelengths and the high precision of the full integration at longer wavelengths.
As the steps of the late-ramp-fit occur prior to light curve fitting, it is computationally inexpensive to perform given the benefit it provides for observations with few groups per integration.
Furthermore, as we show in Appendix \ref{app:nirspec}, the presence of charge migration in other high signal-to-noise near- to mid-infrared instruments like NIRSpec suggests that our methodology will be applicable to affected data sets beyond NIRISS.
We note, however, that the absolute accuracy of the late-ramp-fit cannot be determined from these data alone.
In effect, our method brings the transmission spectrum in the affected wavelength region closer to the more-linear range of the NIRISS instrument's HgCdTe pixels within read pair 1.
We expect that this should improve its accuracy, but a test of this methodology on a transmission spectrum that is known to a high degree of certainty to be flat and featureless will be necessary to confirm it.

\section{JWST NIRISS Reduction} \label{sec:reduction}
We now apply our late-ramp-fit method among the other steps in the \texttt{jwst} pipeline to obtain the NIRISS transmission spectrum of KELT-7\,b.

\subsection{Observations}

The KELT-7 system was observed by JWST's NIRISS instrument using the Single Object Slitless Spectroscopy \citep[SOSS;][]{NIRISS4} mode between 2025 February 15 21:16:51 UTC and 2025 February 16 07:46:35 UTC as part of JWST GO Program 5924 (PI: D. Sing), totaling 9.84 hours.
This observation used the GR700XD grism with the CLEAR filter, Substrip 96 subarray, and NISRAPID readout mode, observing a single transit of KELT-7\,b across the wavelength range 0.85 $\mu\text{m}$--2.85 $\mu\text{m}$.
The entire exposure totaled 3045 integrations at three groups per integration, covering the full transit with pre- and post-transit baseline.

\subsection{\firefly}
Our \firefly reduction up to and including the linearity step in Stage 1 is similar to the reduction presented in \citet{Schmidt25} in that we follow the default \texttt{jwst} pipeline and skip the jump step.
We subtract the background by (1) scaling the flux jump caused by the reflection of zodiacal light off the pick-off mirror in the STScI NIRISS SOSS Substrip 96 background model to the flux jump present in the observed data and (2) adding a constant offset to match the flux level of the model SOSS background before and after the jump to the observed data.
We perform this background subtraction in the relatively PSF-free lower left corner of the detector, with a box bounded by spectral pixels 5 \& 400 as well as spacial pixels 5 \& 20.
As we are performing the image cleaning and noise reduction steps for each group due to our late ramp fit, we perform a group-level $1/f$ subtraction only following the method presented in recent \firefly NIRISS reductions \citep[e.g.,][]{Liu25, Schmidt25, Wang26}.
We clean bad pixels by flagging ones with sharp variance spikes of over 100$\sigma$ using \texttt{lacosmic} \citep{lacosmic}.
We then align frames and perform spectral box extractions for each group using a 31-pixel width aperture and an offset of 10 pixels from the \texttt{pastasoss} \citep{pastasoss2, pastasoss1}-derived trace for Substrip 96 data\footnote{This is a known issue with \texttt{pastasoss}.}.
We select this aperture width as it minimizes the scatter in the white light curve.
Using the late-ramp-fit method described in Section \ref{sec:chargemigration}, we then perform the ramp fit and proceed to spectroscopic light curve fitting.

\begin{figure}
    \centering
    \includegraphics[width=\linewidth]{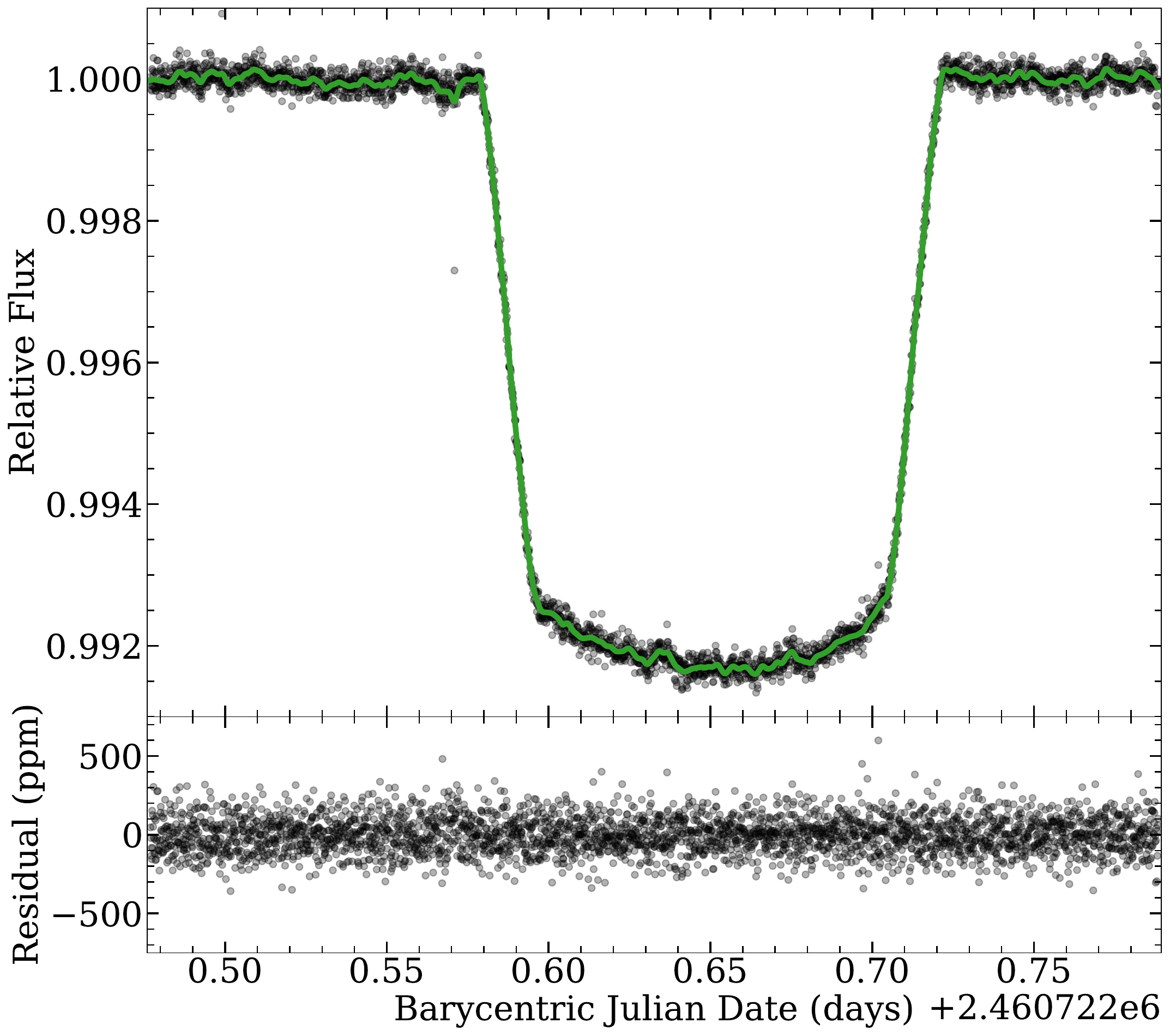}
    \caption{\firefly white light curve of the NIRISS transit of KELT-7\,b. In the upper panel, we plot as black points the individual integration-level measurements, and as the green line our median transit model after exploring the parameter space with \texttt{emcee}. This model includes a linear systematic trend in time and a common-mode correction to mitigate red noise in both the white light curve and transmission spectrum fit. In the lower panel, we plot the residuals between the data and the transit model.}
    \label{fig:wlc}
\end{figure}

Our \firefly white light curve fit, which we show in Figure \ref{fig:wlc}, is based on a \texttt{batman} transit model \citep{Kreidberg2015} and utilizes the Markov Chain Monte Carlo (MCMC) algorithm provided by the \texttt{emcee} Python package \citep{Foremak-Mackey2013}, with 2000 iterations following a burn-in of 200 iterations.
Starting with the raw spectrophotometric time series, we first trim the first and last ten pixel columns of our spectrum.
We choose a systematics vector composed of a Bayesian Information Criterion-preferred linear term in time as well as a common-mode correction, which we obtain by applying an eight integration-wide Gaussian filter on the residual of a white light curve fit whose model's systematics vector consists solely of the linear trend in time; this empirical filter choice captures the correlated noise signals well.
We fit for the baseline flux, planet-to-star radius ratio R$_{\text{p}}/$R$_\ast$, the scaled semimajor axis a/R$_\ast$, the impact parameter b, the time of mid-transit T$_0 - 2460722$, quadratic limb darkening coefficients $u_{+} = u_1 + u_2$ and $u_{-} = u_1 - u_2$, and the systematic vector terms.
We report the orbital parameters for our white light curve fit in Table \ref{tab:orbitparams}.

For our spectroscopic light curve fit, we fix the orbital parameters to the fitted white light curve values and bin the data to $R\sim200$, totaling 238 spectroscopic bins. 
To minimize the additional scatter caused by fitting for limb darkening in this step, we perform a two-step process of (1) fitting for spectroscopic limb darkening coefficients, and (2) fixing them to an offset stellar atmosphere model.
Using the quadratic $u_{+} = u_1 + u_2$ and $u_{-} = u_1 - u_2$ parameterization, we calculate the $\chi^2$ between the spectroscopic theoretical and measured limb darkening coefficients with a constant offset for each parameter across several interpolated stellar atmosphere models using the \texttt{ExoTiC-LD} Python package \citep{grant2024exoticLDJoss}, assuming the stellar atmospheric metallicity, effective temperature, and surface gravity we present in Section \ref{sec:stellar}, choosing the non-density cut parameters to avoid using parameters informed by this transit fit.
We find that these fitted limb-darkening coefficients are best reproduced by the limb darkening coefficients from the \texttt{MPS-ATLAS} model set 1 grid of stellar atmosphere models \citep{Kostogryz22}, setting the minimum $\mu$ value to 0.2 following \citet{Sing10} and with a wavelength-independent offset of $0.06031$ for $u_{+}$ and $-0.28518$ for $u_{-}$.
We use these offset theoretical limb darkening coefficients in our spectroscopic light curve fits.
We fit for the planet-to-star radius ratio $R_{\text{p}}/R_\ast$, baseline flux, and systematic terms in each bin using the Levenberg-Marquardt algorithm provided by the \texttt{lmfit} Python package \citep{lmfit}. 
We account for red noise in our spectrophotometric fit by inflating error bars in quadrature for wavelength bins where the residual RMS-bin size trend is above log linear \citep{Pont2006, Winn2008}.  
As the short-wavelength data is most affected by non-wavelength dependent red noise, our common-mode correction shrinks these points' error bars by roughly a factor of two.
We additionally searched for limb asymmetries by repeating our light curve fitting procedure using the \texttt{catwoman} Python package \citep{Espinoza21, Jones22}, which we show in Figure \ref{fig:catwoman}, but we do not find any significant deviation from the spherical transmission spectrum.

\begin{figure*}
    \centering
    \includegraphics[width=\linewidth]{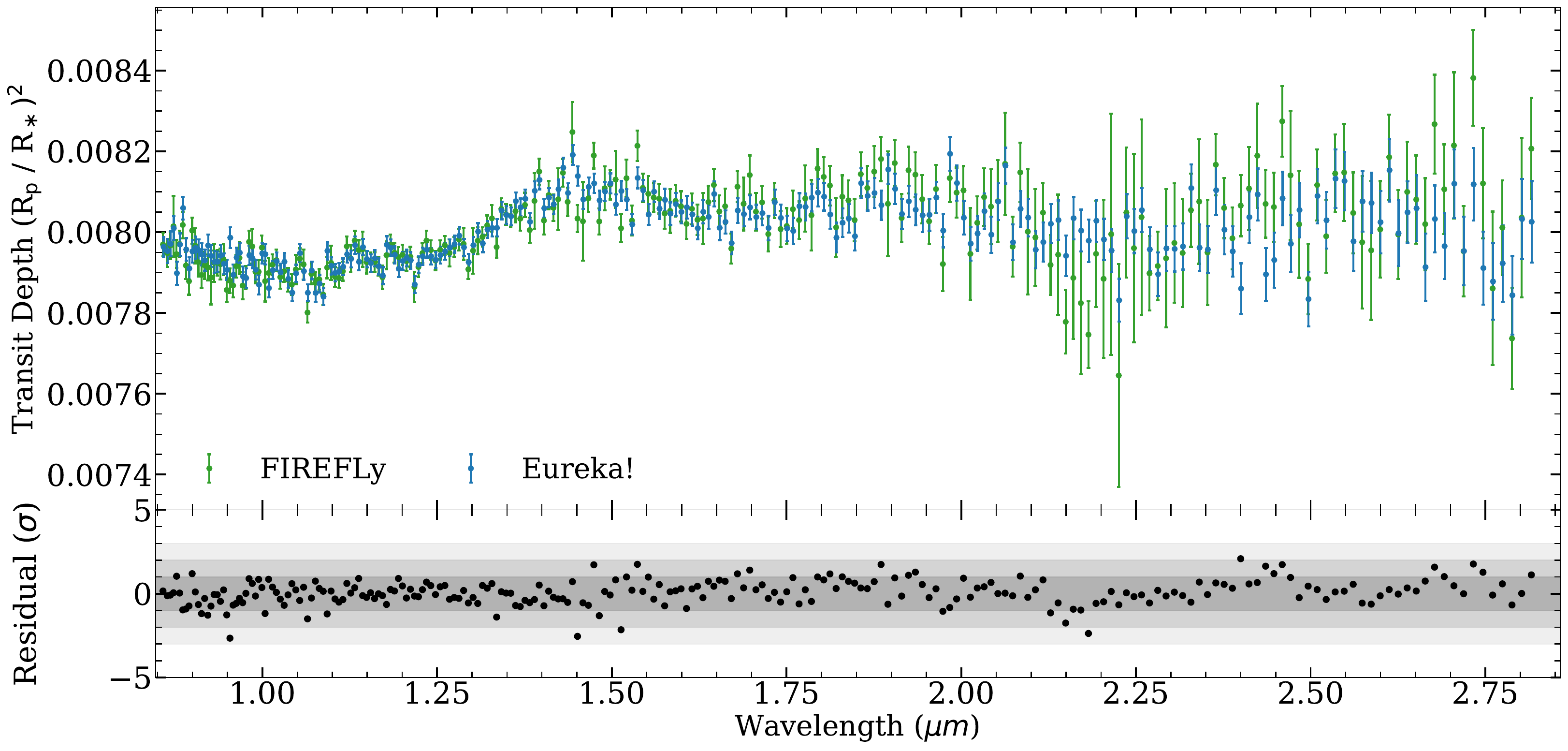}
    \caption{The reduced NIRISS transmission spectrum of KELT-7\,b at $R\sim200$ resolution. In the top panel we plot as green points the \firefly reduction and as blue points the \eureka reduction. 
    Both reductions perform spectral extractions for each group prior to fitting the ramp, a technique to mitigate vertical charge migration-caused nonlinearity we call the ``late-ramp-fit'' method. 
    The error bars for the \firefly reduction are slightly larger than the error bars for the \eureka reduction as \firefly inflates error bars to account for correlated noise.
    We apply a 44 ppm offset to the \eureka reduction to account for our application of a common-mode correction in our \firefly reduction.
    In the lower panel we show the residual between the reductions.
    We also plot as gray regions with increasing transparency the 1-$\sigma$, 2-$\sigma$, and 3-$\sigma$ ranges.
    Our reductions show point-by-point agreement across NIRISS/SOSS Order 1's wavelength range, demonstrating that our late-ramp-fit methodology yields consistent results despite differing reduction choices between the two independent pipelines.}
    \label{fig:spectrum}
\end{figure*}

\subsection{\eureka}
We also use the \eureka pipeline to perform a reduction of the NIRISS/SOSS observations using the late-ramp-fit methodology.
The first two stages of \eureka serve as a wrapper for the standard \texttt{jwst} pipeline, with options for additional data cleaning at the group level.
In this reduction, in Stage 1 we extract read pairs rather than fitting for the ramp as is standard. 
To do this, we mask all but the two groups being considered (on one group, for the first group/read pair, turning the \texttt{suppress\textunderscore one\textunderscore group} flag to \texttt{False} so that the single group is not ignored). 
We otherwise follow the standard Stage 1 and Stage 2 JWST pipeline reduction steps.

\eureka Stage 3 performs 1D time series spectral extraction using the \citet{horne1986optimalE} optimal spectral extraction method. 
Prior to extracting the spectrum we perform a double-iteration outlier identification in the background region along the time axis with an outlier threshold of 5$\sigma$.
We then perform a column-by-column background removal in a region 20 pixels away from the center of the trace using an outlier threshold of 25 $\sigma$ and a flat (i.e 0$^{th}$ order) background level. 
We set the spectral aperture to a half width of 18 pixels, not including the central pixel, for a total of 37 pixels.
We create a median optimal profile using a 5-$\sigma$ outlier threshold. 
During spectral extraction we apply a 7-$\sigma$ outlier threshold when comparing to the median frame. 
We determined all outlier thresholds and aperture sizes to optimize the noise level of the white light curve. 

\begin{figure*}
    \centering
    \includegraphics[width=\linewidth]{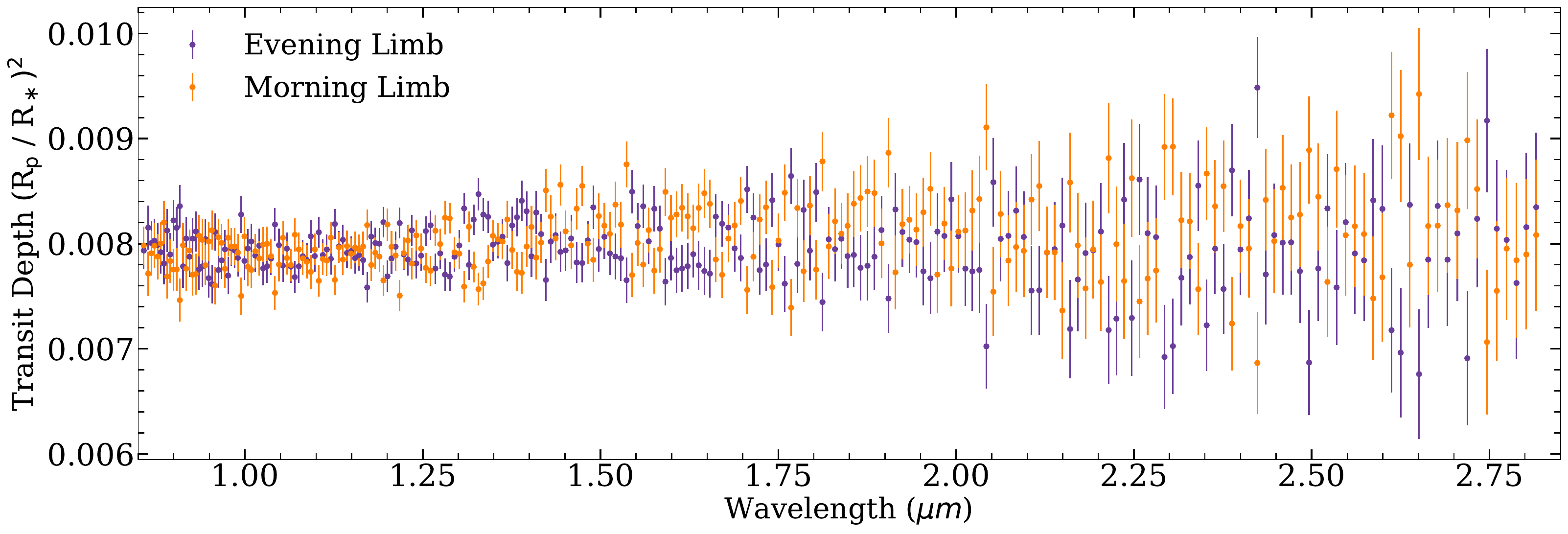}
    \caption{Limb-resolved \texttt{catwoman} reduction of KELT-7\,b's NIRISS transmission spectrum. We plot as the purple error bars the evening terminator's transmission spectrum and as the orange error bars the morning terminator's transmission spectrum. The morning and evening limb transit depths are strongly anti-correlated and are consistent with each other across almost all of NIRISS SOSS's wavelength range, indicating that KELT-7\,b's atmosphere does not exhibit limb asymmetry.}
    \label{fig:catwoman}
\end{figure*}

In Stage 4 of the \eureka pipeline, we bin the 1D time series spectra spectrally into light curves from 0.9 to 2.8 $\micron$. 
Here we use the same binning scheme as \firefly in addition to a single white light curve spanning the full range. 
We calculate limb darkening parameters in this stage using the \texttt{MPS-ATLAS} model set 2 \citep{Kostogryz2023mpsatlas2} stellar atmosphere grid, assuming [M/H] = 0.0, T$_{eff}$ = 6767 K, and log(g) = 4.1 (cgs) \citep{Stassun2019}.
Stage 5 of the \eureka pipeline performs light curve fitting. 
We use \texttt{batman} for the astrophysical signal with quadratic limb darkening and a linear polynomial in time.
For the white light curve fit, we fix the limb darkening to the previously calculated values and fit for a linear polynomial in time, center of transit time $T_0$, scaled semimajor axis a/R$_{s}$, inclination $i$, and transit depth $R_{\text{p}}/R_\ast$ while holding the orbital period constant to the literature value of 2.73477 days \citep{Stassun2017}. 
We report the best-fit values for this fit in Table \ref{tab:orbitparams}. 
For the spectroscopic light curves we again fix the limb darkening to the previously calculated values and hold the orbital parameters constant at the values determined from the white light curve, fitting only for the transit depth and linear polynomial in time.
We use \texttt{emcee} to perform these fits with 250 walkers, 500 burn-in steps, and 2000 steps in the chain. 
We attribute the different fit orbital parameters in our \eureka reduction compared to \firefly to our limb darkening choice; nevertheless, our spectra are in excellent agreement, with the vast majority of points consistent to within 1-$\sigma$.

\begin{deluxetable}{lcc}
\centering
\tablewidth{0pt}
\tablecaption{NIRISS White Light Curve Orbital Parameters}
\tablehead{
Model Setting & \colhead{\firefly} & \colhead{\eureka}
}
\startdata
T$_0$ - 2460722 & $0.650199 \pm 0.000003$ & $0.650168 \pm 0.000017$ \\
a/R$_\ast$ & $5.586 \pm 0.002 $ & $5.545 \pm 0.012$\\
b & $0.594 \pm 0.001$ & $0.604 \pm 0.004$ \\
i & $83.90\pm0.01$ & $83.75 \pm 0.04$ \\
$\left(\text{R}_{\text{p}}/\text{R}_\ast\right)^2$ & $0.007981 \pm 0.000003$  & $0.007965 \pm 0.000006$ \\
\enddata
\tablecomments{$T_0$ is the time of transit in BJD, $a/R_\ast$ is the unitless scaled semimajor axis, $b$ is the unitless impact parameter, $i$ is the inclination in degrees, and $\left(R_p/R_\ast\right)^2$ is the squared planet-to-star radius ratio. As \firefly and \eureka use different parameters to describe the inclination (impact parameter for \firefly and inclination for \eureka), we also calculate and report the unfit-for parameter for each pipeline.}
\label{tab:orbitparams}
\end{deluxetable}

We show our resulting transmission spectra in Figure \ref{fig:spectrum}, and they are available for download in this article's \href{https://www.zenodo.org/records/20419021}{Zenodo repository}.

\section{HST UVIS Re-reduction} \label{sec:hst}

Inclusion of HST WFC3/UVIS G280 data in our free retrievals would help constrain inhomogeneities on KELT-7's surface as well as the presence of aerosols or refractory species like TiO and VO in KELT-7\,b's atmosphere.
However, the high resolution of the HST WFC3/UVIS G280 spectrum presented in \citet{Gascon25} tends to lead to overfitting of stellar contamination in our free retrievals.
To mitigate this, we re-reduce these observations using a modified version of the \firefly pipeline that has been used in recent UVIS G280 analyses \citep[e.g.,][]{Bennett25}.
Starting with the raw \texttt{.flt} files from MAST, we perform a background subtraction using the most up-to-date background data from STScI.
We then perform cosmic ray removal using \texttt{lacosmic}.
We calculate the wavelength solution and trace position based on the UVIS G280 calibration presented in \citet{Pirzkal20}.
We measure X-Y shifts by calculating the peak of the cross-correlation between each frame and the first frame in the series.
We then extract spectra for both the +1 and -1 orders by selecting a box corresponding to the minimum and maximum X and Y bounds of the trace.
For order +1, we add an additional 10 pixels above and below this box to capture the full trace; for order -1, we instead add 7 pixels above and below due to its different shape.
Finally, we sum each column to obtain the transmission spectrum.

Following a similar base transit model as our JWST reduction, we fit the light curves for each order using \texttt{emcee}, fixing the orbital parameters to the values we obtained from our NIRISS reduction.
After including the detrending vectors from the observation's \texttt{.jit} files, we choose a systematics vector for each order based on the Akaike Information Criterion (AIC).
For order +1 this includes \texttt{V2\_dom}, \texttt{V2\_roll}, \texttt{V3\_roll}, \texttt{RA}, \texttt{Dec}, \texttt{LimbAng}, \texttt{Lat}, X-shifts, Y-shifts, and a linear trend in time.
For order -1 this includes \texttt{V3\_dom}, \texttt{V2\_roll}, \texttt{RA}, \texttt{Dec}, \texttt{LimbAng}, \texttt{Lat},  \texttt{Long}, \texttt{lin\_orb}, X-shifts, and a linear term in time\footnote{More details about these systematic vectors are available in \citet{Sing19}.}.
In our spectroscopic fits we first bin the data to $R\approx10$.
We then use the same limb darkening prescription as NIRISS, including the parameterization, stellar atmosphere model used, and wavelength-independent offsets for each limb darkening parameter.
Constant offsets between calculated limb darkening coefficients and theoretical limb darkening coefficients have been shown to improve pipeline scatter for transmission spectrum observations \citep[e.g.,][]{Rus23, Sing24a}. 
They have also been shown to be a good approximation for the true stellar limb darkening coefficients among the population of transiting giant exoplanets \citep{Sing2026}, hence our extrapolation here.
We fit each bin using \texttt{lmfit}, fixing the systematics vector to the white light curve values.
We then average the +1 and -1 orders to obtain the final transmission spectrum.

\begin{figure*}
    \centering
    \includegraphics[width=\linewidth]{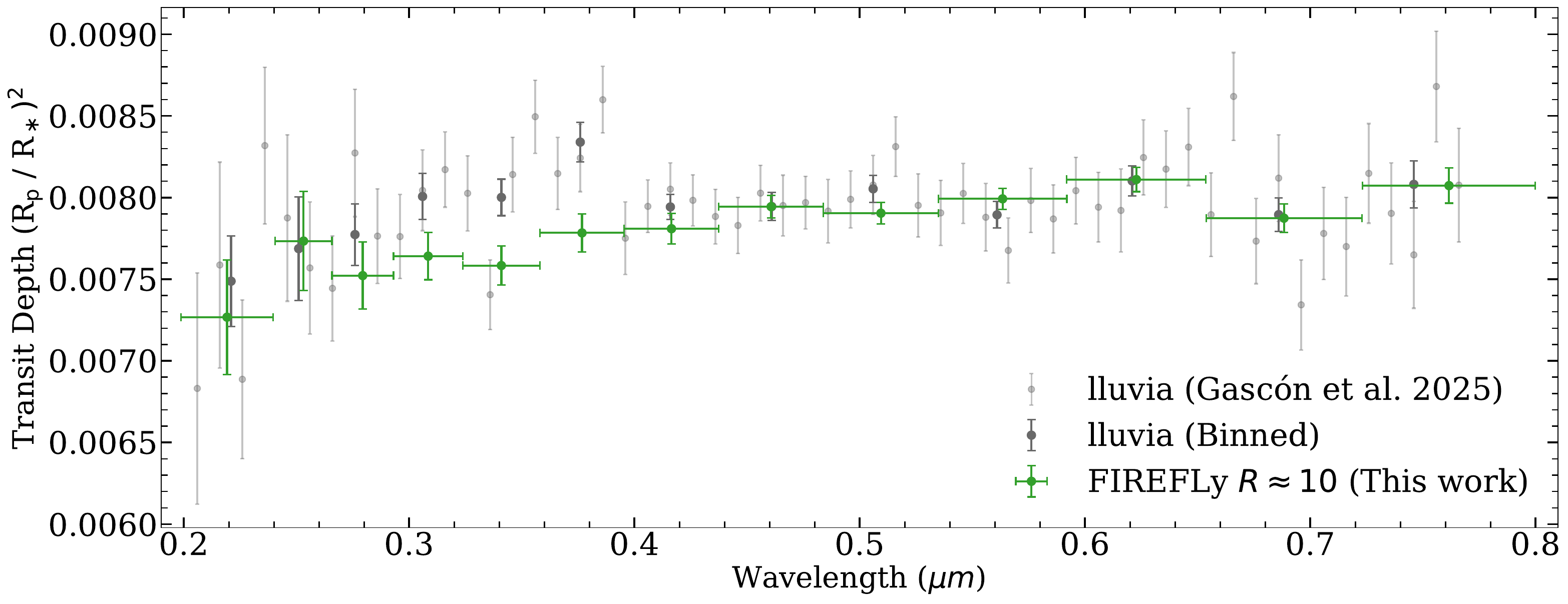}
    \caption{Comparison between our re-reduced HST WFC3/UVIS G280 transmission spectrum and archival higher-resolution data. Our reduction, which we plot as the green points, was performed using a modified version of the \firefly pipeline. For comparison, we plot as gray points the \texttt{lluvia} combined-order transmission spectrum from \citet{Gascon25} at the original resolution (small points, transparent) and binned to \firefly's resolution (large points, opaque). We prefer to use lower resolution HST UVIS data in our \poseidon retrievals to avoid overfitting stellar contamination. The consistency between the archival data and our reduction demonstrates the fidelity of the \firefly pipeline for HST WFC3/UVIS G280 observations.}
    \label{fig:uvis}
\end{figure*}

We show the results of this reduction in Figure \ref{fig:uvis}. 
Our reduction is at lower resolution than the spectrum presented in \citet{Gascon25}, though their shapes are consistent with one another without the need for a wavelength-independent shift in transit depth.
We proceed to use this HST transmission spectrum in our free retrieval analysis.

\section{Stellar Characterization} \label{sec:stellar}
In order to accurately infer the atmospheric and fundamental properties of KELT-7\,b, we also require precise and accurate stellar properties.
The precise parallax measurements presented in Gaia Data Release (DR) 3 significantly improve the precision of stellar parameters inferred by fitting multi-wavelength photometry and astrometry to theoretical stellar evolution models; however, to date no detailed characterization of the host star KELT-7 has been presented in the literature since the release of Gaia DR2 or DR3. 
We therefore use the \texttt{Starlord} Python package\footnote{\url{www.github.com/dpthorngren/starlord}} with archival stellar photometric data, Gaia DR3 astrometry, and three-dimensional local extinction and reddening maps to infer accurate and precise fundamental and photospheric stellar parameters of KELT-7. 
\texttt{Starlord} performs simultaneous Bayesian fits of the Modules for Experiments in Stellar Evolution \citep[MESA;][]{pax11,pax13,pax18,pax19,jer23} Isochrones \& Stellar Tracks \citep[MIST;][]{dot16,cho16} isochrone grid to a curated collection of data for the star using an MCMC simulation. 
This methodology, with heritage in our previous use of the \texttt{isochrones} Python package to perform these fits \citep{mor15}, has been extensively used in the literature \citep[e.g.,][McCreery et al. in preparation]{Hamer22, Schmidt23, Schmidt26b} and has become the state-of-the-art when inferring stellar parameters. 
We use the derived stellar mass and radius of KELT-7 to then update the planet properties of KELT-7\,b that we use in our retrieval analysis.

The precise photometry from the JWST transit light curve of KELT-7\,b enables our use of the stellar mean density as an additional constraint on KELT-7's fundamental properties.
From Kepler's third law, it is possible to directly relate the stellar mean density to the scaled semimajor axis $a/R_\ast$, allowing a model-independent constraint on the host star's structure \citep{Sozzetti07}.
This has been demonstrated for other transiting systems observed by JWST as a means of obtaining highly precise stellar properties \citep[e.g.,][Rustamkulov et al., under review]{Sing24a}.
We first select the $a/R_\ast$ and uncertainty from our \eureka reduction, the \eureka NRS1 \& NRS2 joint fit $a/R_\ast$ from \citet{Ahrer25}, and the TESS-derived $a/R_\ast$ from \citet{Patel22}.
We use \eureka in both JWST cases to limit pipeline-dependent systematics.
We calculate the variance-weighted mean $a/R_\ast$, adding in quadrature to each uncertainty the standard deviation of these measurements as a constant systematic uncertainty.
We then compute the stellar mean density following \citet{Seager03} (note a dropped factor of $3/4\pi$ prior to Equation 19):
\begin{equation}
    \rho_{\ast, \text{transit}} = \frac{3\pi}{GP^2}\left(\frac{a}{R_\ast}\right)^3.
\end{equation}
With this, we infer a stellar mean density $\rho_{\ast,\text{transit}} = 0.433 \pm 0.005$ g cm$^{-3}$. 
This transit-inferred mean density, in conjunction with broad-band photometry, enables the precise calculation of KELT-7's mass, age, and metallicity.

We fit the MIST grid to the following data:
\begin{enumerate}
\item Galaxy Evolution Explorer \citep[GALEX;][]{mart05} near-ultraviolet (NUV) photometry from the GUVcat\_AIS \citep{bia17} including in quadrature a zero-point uncertainty of 0.02 mag;
\item Tycho-2 $B_{T}$ and $V_{T}$ photometry including in quadrature their zero-point uncertainties (0.078, 0.058) mag \citep{hog00,marz05};
\item Gaia EDR3 $G$ photometry including in quadrature its zero-point uncertainty \citep{Gaia_mission2016, GaiaEDR3, GaiaEDR3Validation, EDR3photometry, row21, tor21};
\item Two-micron All-sky Survey (2MASS) $JHK_{s}$ photometry including their zero-point uncertainties \citep{skr06};
\item Wide-field Infrared Survey Explorer (WISE) All-sky \citep{wri10} $W1$ and $W2$ photometry in quadrature their zero-point uncertainties (0.032,0.037) mag\footnote{\url{https://wise2.ipac.caltech.edu/docs/release/allsky/expsup/sec4\_4h.html\#PhotometricZP}}.
\item We also fit to a zero point-corrected Gaia EDR3 parallax \citep{GaiaEDR3, GaiaEDR3Validation, lin21a, lin21b, row21, tor21},
\item An estimated extinction value based on the 3D local extinction maps using the \texttt{G-Tomo} Python module \citep{lal22, ver22}, and
\item Our transit-informed stellar mean density.
\end{enumerate}
We use the data quality flags described in \citet{Hamer22}.  
We do not include the absolutely-calibrated JWST stellar spectrum in this analysis as it does not significantly improve our result.

For our priors, we use a \citet{cha03} log-normal mass prior for $M_{\ast} < 1~M_{\odot}$ joined to a \citet{sal55} power-law prior for $M_{\ast} \geq 1~M_{\odot}$, a metallicity prior based on the Geneva-Copenhagen Survey \citep[GCS;][]{cas11}, a log-uniform age prior between 100 Myr and the age of the universe 13.721 Gyr, a uniform extinction prior in the interval between the value inferred by 3D local extinction and reddening maps minus/plus five times its uncertainty, and a distance prior proportional to volume between the \citet{bai21} geometric distance minus/plus five times its uncertainty.

We report the output of our \texttt{Starlord} Bayesian inference in Table \ref{tab:stellarprops}.
Both the mean density and GALEX ultraviolet photometry \citep[Which strongly constrains the metallicity of the star;][]{Nataf2024} independently move the stellar posterior in the direction of a younger, more massive, and metal-rich star.
This results in a star more metal-enriched and younger than a less comprehensive inference, such as the first analysis of KELT-7 presented by \citet{Bieryla15}, would suggest.
As they performed their analysis prior to the first Gaia data release, \citet{Bieryla15} were unable to incorporate photometric measurements into their fit as they lacked a precise parallax measurement.
As a result, they were restricted to inferring the atmospheric metallicity of KELT-7 via its (rotationally-broadened) TRES spectrum.
Our fit, on the other hand, incorporates KELT-7's photometry, precise Gaia astrometry, and stellar density, and is not affected by rotational broadening.
This explains the discrepancy between our inferred metallicity and the metallicity found by \citet{Bieryla15}.

\begin{figure*}
    \centering
    \includegraphics[width=.85\linewidth]{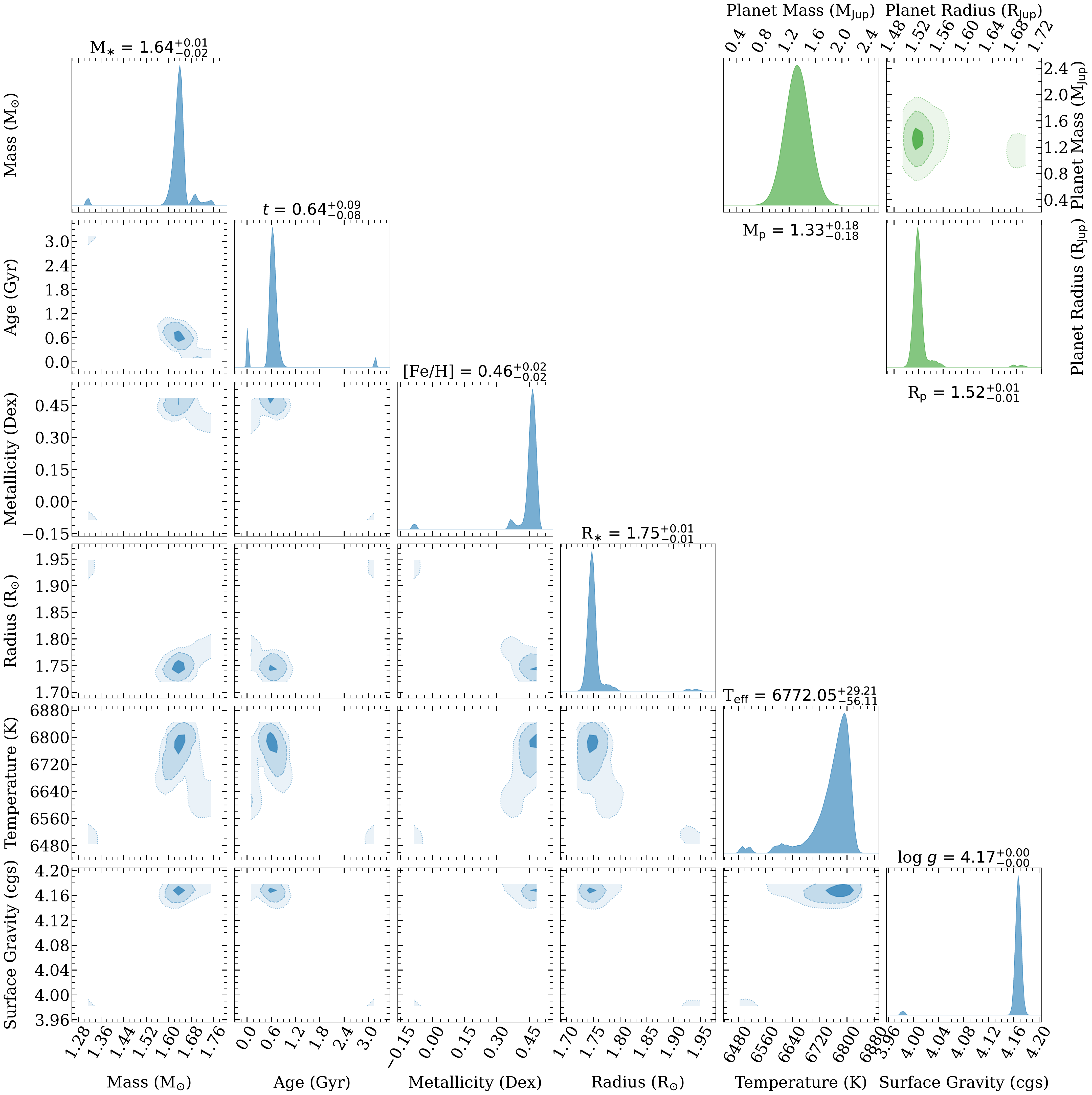}
    \caption{Corner plot showing the posterior distribution of KELT-7's fundamental stellar parameters (blue) and subsequent inference of KELT-7\,b's mass and radius (green).}
    \label{fig:stellar}
\end{figure*}

Using the Doppler semiamplitude reported by \citet{Bieryla15} and NIRISS white light curve planet-to-star radius ratio with our updated stellar mass and radius, we follow the procedure of \citet{Schmidt23} to calculate an updated mass and radius for KELT-7\,b, which we report in Table \ref{tab:stellarprops}.
We also calculate an updated planetary surface gravity, $\log g = 3.15 \pm 0.07$, and zero-albedo equilibrium temperature, T$_{\text{eq}}=2025$ K.
Using these updated stellar and planetary properties, we proceed to perform a retrieval analysis on KELT-7\,b's panchromatic transmission spectrum.

\begin{deluxetable*}{lccc}\label{tab:stellarprops}
    \centering
    \tablewidth{0pt}
    \tablecaption{KELT-7 System Properties}
    \tablehead{Parameter & Description & No Density Cut/GALEX & With Density Cut/GALEX}
    \startdata
    \hline
    \multicolumn{4}{c}{Stellar Properties}\\
    \hline
    M$_\ast$ & Mass (M$_\odot$) & $1.43^{+0.09}_{-0.06}$ & $1.64^{+0.02}_{-0.02}$ \\
    $t_\ast$ & Age (Gyr) & $1.97^{+0.37}_{-0.51}$ & $0.64^{+0.10}_{-0.09}$  \\
    $[\text{Fe}/\text{H}]$ & Photospheric Metallicity (dex) & $0.07^{+0.21}_{-0.18}$ & $0.46^{+0.02}_{-0.02}$ \\
    R$_\ast$ & Radius (R$_\odot$) & $1.79^{+0.02}_{-0.02}$ & $1.75^{+0.01}_{-0.01}$ \\
    T$_{\text{eff}}$ & Effective Temperature (K) & $6630^{+60}_{-50}$ & $6770^{+30}_{-60}$ \\
    $\log g_\ast$ & Surface Gravity (cgs) & $4.09^{+0.03}_{-0.03}$ & $4.17^{+0.01}_{-0.01}$ \\
    \hline
    \multicolumn{4}{c}{Planet Properties}\\
    \hline
    M$_{\text{p}}$ & Mass (M$_{\text{Jup}}$) & $1.22^{+0.18}_{-0.17}$ & $1.33^{+0.18}_{-0.18}$ \\
    R$_{\text{p}}$ & Radius (R$_{\text{Jup}}$) & $1.56^{+0.02}_{-0.02}$ & $1.52^{+0.01}_{-0.01}$ \\
    $\log g_{\text{p}}$ & Surface Gravity (cgs) & $3.09^{+0.06}_{-0.06}$ & $3.15^{+0.07}_{-0.07}$\\
    T$_{\text{eq}}$ & Zero-albedo Equilibrium Temperature (K) & 1984 & 2025 \\
    \enddata
\end{deluxetable*}

\section{Retrieval Analysis} \label{sec:retrievals}
We conduct a retrieval analysis on the combination of the JWST NIRISS transmission spectrum calculated in Section \ref{sec:reduction}, the JWST NIRSpec/G395H transmission spectrum presented in \citet{Ahrer25}, and the HST WFC3/UVIS G280 transmission spectrum re-reduced in Section \ref{sec:hst}.
While our free retrieval considers all of HST WFC3 UVIS/G280, JWST NIRISS/SOSS, and JWST NIRSpec/G395H, each of our equilibrium chemistry models only consider NIRISS and NIRSpec data.
As the wavelength range of the HST WFC3/IR G141 transmission spectrum presented in \citet{Pluriel20} overlaps in its wavelength range with our NIRISS transmission spectrum, we do not include it in our retrieval analyses.
We perform our retrieval analysis using two open-source retrieval codes, \poseidon and \pRT, and we summarize the configuration for each in Table \ref{tab:retrieval_configurations} in Appendix \ref{app:opacities}.
In addition, we perform a grid fit using the \picaso planetary atmosphere modeling code to contextualize these results against a physically self-consistent equilibrium chemistry model; we summarize its configuration in Table \ref{tab:picaso}.

\subsection{\poseidon} \label{sec:POSEIDON}

We use the \poseidon Python package \citep{MacDonald2017,MacDonald2023} to perform both free chemistry and equilibrium chemistry atmosphere retrievals on KELT-7\,b's transmission spectrum.
Our atmosphere model assumes background gas consisting of H$_2$ and He at Solar He/H$_2$ = 0.17 \citep{asplund09}.
We construct atmospheres with 150 layers uniformly spaced in log pressure, ranging from 10$^{-7}$-10$^2$ bar, under hydrostatic equilibrium.
We solve for hydrostatic equilibrium with the assumption that that the atmosphere has a pressure of 10$^{-3}$ bar at the reference radius.
We freely retrieve the six $T(P)$ profile parameters following the prescription of \citet{Madhusudhan2009}.
Our equilibrium chemistry atmosphere model considers the following trace species: H$_2$O, CH$_4$, CO$_2$, CO, HCN, SO$_2$, H$_2$S, OCS, TiO, VO, and SiO; our free chemistry model also includes H$^-$ bound-free absorption and additional high-temperature species AlO, SH, and FeH\footnote{We do not include these in the equilibrium chemistry \poseidon retrieval as its \texttt{FASTCHEM} grid does not include them.}.
We retrieve the log volume mixing ratios (VMRs) for each of the trace species in our model for our free chemistry retrievals.
For our equilibrium chemistry retrievals, we instead retrieve the atmospheric metallicity and C/O ratio as calculated from the \poseidon package's \texttt{FASTCHEM} grid \citep{FASTCHEM}.
For clouds and aerosols, we use the ``slab'' Mie scattering prescription presented in \citet{Mullens24}, where we vary the log volume mixing ratio of corundum Al$_2$O$_3$, the only aerosol that may be condensed at the temperature of KELT-7\,b's atmosphere \citep{Sing13, Wakeford15}, the particle size, the cloud top pressure of the cloud slab, and the pressure range of the cloud slab.
We use the opacity data for the combination of amorphous alumina and $\gamma$ crystalline corundum \citep[i.e., the \texttt{Al2O3\_KH} species in \poseidon;][]{Kitzmann18, Begemann97, Koike95} to allow more of our re-reduced WFC3/UVIS spectrum to be used than with pure crystalline corundum\footnote{At the wavelengths we are considering, the difference in resulting transmission spectra between it and pure $\gamma$ crystalline corundum is negligible.}.
As our free retrieval covers shorter wavelengths than our equilibrium retrieval, we marginalize over possible stellar contamination from unocculted starspots or faculae by interpolating PHOENIX stellar atmosphere models \citep{husser2013phoenix} using the \texttt{PyMSG} package \citep{Townsend2023} and the three-parameter stellar contamination prescription from \citet{Rathcke2021}.
We allow for three free offsets, one between the NIRISS/SOSS and NIRSpec/G395H NRS1 data, another between the NIRISS/SOSS and NIRSpec/G395H NRS2 data, and a third between the NIRISS/SOSS and WFC3/UVIS G280 data (when it is being considered).
In total, our equilibrium chemistry model has 15 free parameters: the reference radius, 6 $T(P)$ profile parameters, 4 aerosol parameters, 2 equilibrium chemistry parameters, and offsets between NIRISS/SOSS and the two detectors of NIRSpec/G395H.
Our free chemistry model also includes 3 stellar contamination parameters, an offset between NIRISS/SOSS and HST WFC3/UVIS, and replaces the metallicity and C/O ratio with VMRs for 15 trace species, totaling 32 free parameters.

We calculate model spectra in our \poseidon retrievals by solving the equation of radiative transfer in a cylindrical coordinate system for 100 incident stellar rays that are attenuated according to the atmospheric opacity along line-of-sight.
We pre-compute our opacities at $R=20,000$ across a grid of temperatures and pressures using the \texttt{Cthulhu} Python package \citep{Agrawal2024,Cthulu}.
We list our molecular species' opacity sources in Appendix \ref{app:opacities}.
We also include continuum collision-induced absorption from H$_2$-H$_2$ and H$_2$-He pairs \cite{Karman2019} and Rayleigh scattering for all gases \citep{MacDonald2022}. 
In our free retrievals we include H$^-$ bound-free absorption using the line list from \citet{John88}.

In addition to our reference free chemistry model, we perform four nested retrievals: one without stellar contamination, one without the high-temperature species TiO, VO, SiO, AlO, SH, and FeH, and one each without individual species H$_2$O and CO$_2$ to calculate Bayes factors and assess the statistical significance of these components of our model.
We group the high-temperature species into one set of nested retrievals to facilitate a more direct comparison with \citet{Ahrer25}'s test.
We calculate model transmission spectra using opacity sampling on a wavelength grid from 0.2--5.3 $\mu$m (or 0.75--5.3 $\mu$m for our equilibrium retrieval), and we perform nested sampling via the \texttt{MultiNest} algorithm using the \texttt{PyMultiNest} \citep{fer08,fer09,fer19} Python package.
This \texttt{MultiNest} run uses 1000 live points to constrain the parameter space for each of our retrievals.

\subsection{Free Retrieval Results} \label{sec:retrieval_results}
\begin{figure*}
    \centering
    \includegraphics[width=\linewidth]{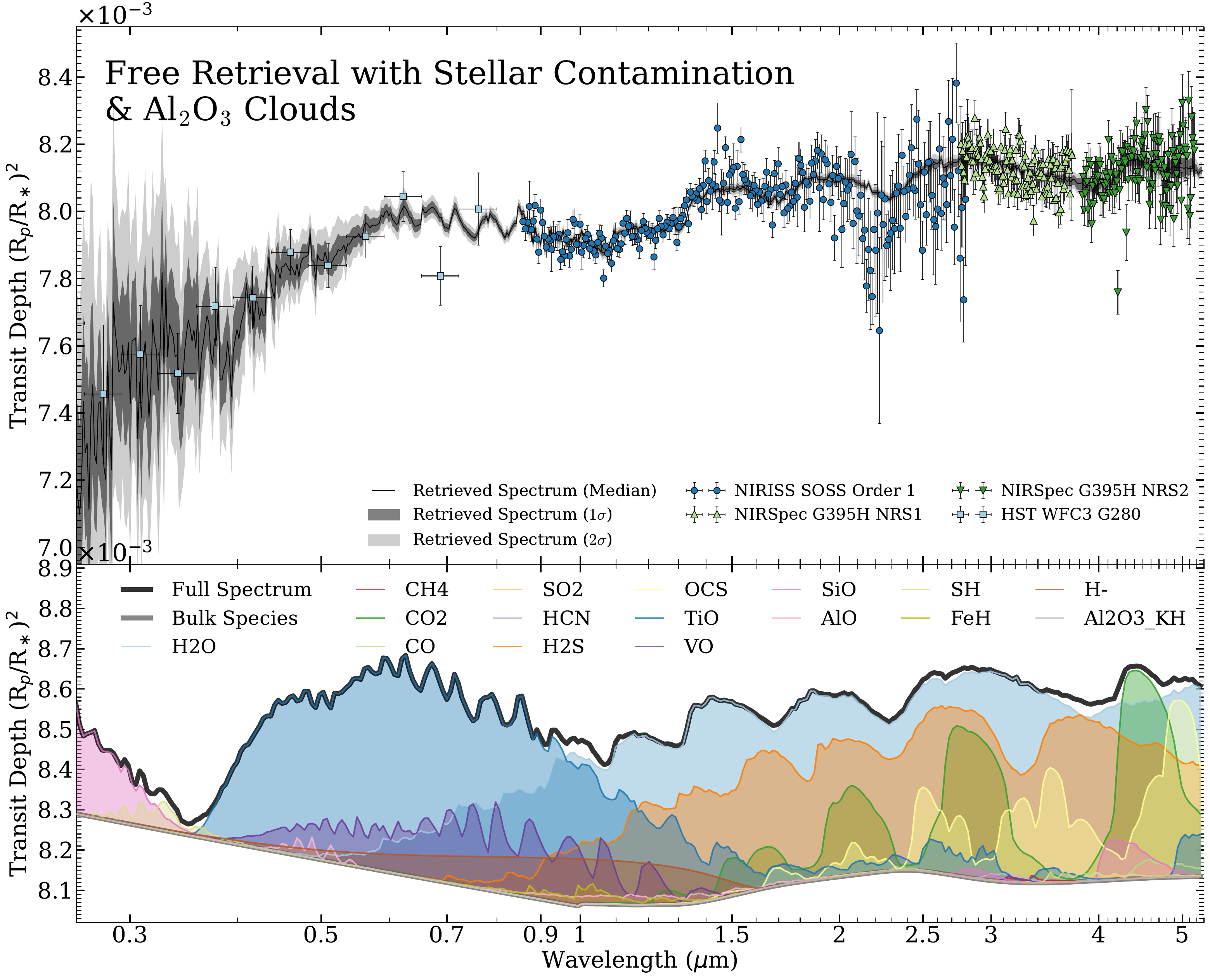}
    \caption{Summary of the results from our \poseidon panchromatic free retrieval. In the upper panel we show a comparison between the retrieved transmission spectrum and the data used in our analysis. 
    We plot as the black line the median retrieved spectrum and as (dark/light) gray polygons the (1/2)-$\sigma$ ranges of the retrieved spectrum.
    We overplot as colored points with black borders the HST WFC3/UVIS (squares), JWST NIRISS (circles), and JWST NIRSpec (triangles) data points, differentiating between data from NIRSpec/G395H NRS1 (light green, upward-facing) and NIRSpec/G395H NRS2 (dark green, downward-facing) by the direction of the triangle as well as the color.
    We apply separate offsets between each detector of the archival JWST NIRSpec/G395H \eureka reduction presented in \citet{Ahrer25} as well as our re-reduced HST WFC3/UVIS G280 transmission spectrum. 
    In the lower panel we show the spectral contribution corresponding to our median retrieved atmosphere.
    We strongly prefer the presence of both stellar contamination ($\log_{10}\mathcal{B}_{\text{contam}} = 8.96$), though we find the fitted values to be unphysical, and the combination of the high-temperature species TiO (dark blue), VO (dark purple), AlO (light pink), SiO (dark pink), SH (light gold), and FeH (darker gold) ($\log_{10}\mathcal{B}_{\text{hi-temp}} = 9.76$).
    We also detect H$_2$O (light blue) and CO$_2$ (dark green) with Bayes factors $\log_{10}\mathcal{B}_{\text{H}_2 \text{O}} = 23.05$ and $\log_{10}\mathcal{B}_{\text{C}\text{O}_2} = 1.82$, indicating decisive evidence for H$_2$O and strong evidence for CO$_2$.
    We do not favor the presence of Al$_2$O$_3$ aerosols (gray) or enhanced H$^-$ opacity (brown) in our retrieval.
    The addition of NIRISS data has enabled a panchromatic study of KELT-7\,b's atmosphere, disentangling questions left open by previous analyses on subsets of these data.}
    \label{fig:poseidon_panchromatic}
\end{figure*}
We show the results of our \poseidon panchromatic free chemistry retrieval in Figure \ref{fig:poseidon_panchromatic}.
Our retrieval exhibits decisive evidence \citep[according to][]{Thorngren26} for the presence of both stellar contamination ($\log_{10}\mathcal{B}_{\text{contam}} = 8.96$) and the combination of the high-temperature species TiO, VO, AlO, SiO, SH, and FeH ($\log_{10}\mathcal{B}_{\text{hi-temp}} = 9.76$).
Our retrieval prefers a stellar contamination fraction $f_{\rm{het}}=0.25^{+0.08}_{-0.07}$, with a heterogeneity temperature $T_{\rm{het}}=6890^{+40}_{-30}$ and photosphere temperature $T_{\rm{phot}}=6700^{+30}_{-40}$; this is driven almost entirely by the lower transit depths of the bluest HST WFC3/UVIS G280 data points and is likely to be at least in part unphysical, perhaps caused by detector edge effects or another systematic.
In particular, we constrain a VMR for TiO of $\log X_{\text{TiO}} = -5.06 ^{+0.21}_{-0.24}$; its presence would cause a thermal inversion in KELT-7\,b's atmosphere and motivates our inclusion of it in our \picaso self-consistent atmosphere models.
We also confirm detections of H$_2$O and CO$_2$ with Bayes factors $\log_{10}\mathcal{B}_{\text{H}_2 \text{O}} = 23.05$ and $\log_{10}\mathcal{B}_{\text{C}\text{O}_2} = 1.82$, indicating decisive evidence for H$_2$O and strong evidence for CO$_2$ in KELT-7\,b's atmosphere at VMRs $\log X_{\text{H}_2\text{O}} = -1.17^{+0.12}_{-0.16}$ and $\log X_{\text{CO}_2} = -3.35^{+0.31}_{-0.38}$.
On the other hand, our \poseidon free retrieval does not detect the presence of CO in KELT-7\,b's atmosphere as it prefers to fit OCS over CO at around 5 $\mu$m. This is implausible under the assumptions of equilibrium chemistry; without a CO detection, the C/O ratio cannot be well constrained in a free retrieval framework.
We also do not detect H$^-$ opacity, in contrast with the previous studies that utilized HST WFC3/IR G141 data, or Al$_2$O$_3$ aerosols in KELT-7\,b's atmosphere.

\subsection{\pRT} \label{sec:pRT}
We perform an independent atmospheric retrieval analysis with \pRT \citep{Molliere19, Nasedkin24} to contextualize our \poseidon equilibrium chemistry results with the NIRSpec analysis presented in \citet{Ahrer25}. 
Because \pRT does not have a built-in treatment of stellar contamination, we also restrict our retrievals to the near-infrared JWST NIRISS/SOSS and NIRSpec/G395H data.
We perform an equilibrium chemistry retrieval with NIRSpec alone as a direct comparison to \citet{Ahrer25} as well as an equilibrium chemistry retrieval with both the NIRISS and NIRSpec data. 
We closely match the setup of the retrievals in \cite{Ahrer25}, modeling an 80-layer atmosphere with a pressure range between $10^3$ and $10^{-6}$ bar.
We assume a 3-parameter \citet{Guillot10} $T(P)$ profile parameterization with an intrinsic temperature $T_{\rm int} = 500$~K, an atmosphere in chemical equilibrium, and an opaque gray cloud deck.
We match the base species included in \citet{Ahrer25} and list our opacity sources in Appendix \ref{app:opacities}. We also tested the use of HITEMP as our source for the H$_2$O line list, which we discuss in Appendix \ref{app:pRTlinelists}.
We included two data offsets: one between NIRISS/SOSS and NIRSpec/G395H NRS1 and another between NIRISS/SOSS and NIRSpec/G395H NRS2. 
We enumerate the full prior constraints and retrieval settings in Table~\ref{tab:retrieval_configurations} in Appendix \ref{app:opacities}. As shown in Appendix~\ref{app:pRTlinelists}, we reproduce the results of \citet{Ahrer25} with an identical retrieval setup, confirming that any differences in the \pRT retrievals when including the NIRISS/SOSS data stem from the additional information it provides rather than differences in the setup itself.

\subsection{\picaso} \label{sec:PICASO}
In addition to our \poseidon and \pRT equilibrium chemistry retrievals with a free $T(P)$ profile, we use the \picaso Python package \citep{Batalha19,Mukherjee23,Mang26}, an open-source\footnote{\url{https://github.com/natashabatalha/picaso}} exoplanet atmosphere model suite adapted from prior planetary thermal structure and atmosphere codes \citep{mckay1989thermal, marley1999thermal, fortney2005comp, cahoy2010exoplanet}, to perform a Bayesian grid fitting analysis of the NIRISS and NIRSpec transmission spectrum of KELT-7\,b.
We generate 100-layer one-dimensional climate models following the procedure presented in \citet{Mukherjee23}, producing homogeneous, physically self-consistent model atmospheres in radiative-convective equilibrium.
Our models incorporate the Sonora grid of equilibrium chemistry models presented in \citet{Marley21}\footnote{\url{https://zenodo.org/records/5063476}; \citep{sonorabobcat}.}.
We list our opacity sources in Appendix \ref{app:opacities}.

For our grid of models, we vary four parameters: the intrinsic temperature T$_{\text{int}}$, the atmospheric metallicity [M/H], the C/O ratio, and the day side-to-night side heat redistribution factor f$_{\text{rfacv}}$ \citep[][Equation 20]{Mukherjee23}.
We select three values of T$_{\text{int}}$: 500 K, 550 K, and 600 K, to explore plausible values for KELT-7\,b as predicted by \citet{Thorngren19}.
To assess the potential effect of increased or inhibited recirculation efficiency, we select 0.3, 0.5, and 0.7 as our grid values for f$_{\text{rfacv}}$.
Here f$_{\text{rfacv}}=0.5$ corresponds to full heat redistribution.
We use the full grid of metallicity and C/O base models with TiO and VO included from the Sonora grid (where available for each combination), as motivated by our free retrieval analysis.
This includes metallicities [M/H] = -1.00 (except C/O = 1.50, 2.00, and 2.50), -0.70, -0.50, -0.30, 0.00, 0.30, 0.50, 0.70, 1.00, 1.30, 1.50, 1.70, and 2.00, as well as C/O ratios 0.25, 0.50, 1.00, 1.50, 2.00, and 2.50.
In total, our grid has 675 individual models.

\begin{deluxetable*}{lccc}\label{tab:picaso}
    \centering
    \tablewidth{0pt}
    \tablecaption{\picaso Grid Fit Priors \& Fitted Values}
    \tablehead{Parameter & Description & Prior & Fit Value}
    \startdata
    \hline
    $\text{T}_{\text{int}}/\text{K}$ & Intrinsic Temperature & $\mathcal{U}(500,600)$ & $552^{+32}_{-33}$ \\
    $\text{[M/H]}$ (dex) & Atmospheric Metallicity & $\mathcal{U}(0,2.5)$ & $1.96^{+0.03}_{-0.27}$ \\
    $\text{C/O}$ (unitless) & Atmospheric C/O Ratio & $\mathcal{U}(0.25,2.5)$ & $0.46^{+0.15}_{-0.13}$\\
    $\text{f}_{\text{rfacv}}$ (unitless) & Recirculation Efficiency & $\mathcal{U}(0.3,0.7)$ & $0.66^{+0.04}_{-0.34}$ \\
    $\log \text{f}_{\text{sed}}$ (unitless) & Sedimentation Factor & $\mathcal{U}(-4,1)$ & $-1.24^{+1.25}_{-1.46}$\\
    $\log \text{K}_{\text{zz}}~\text{(cm$^2$ s$^{-1}$)}$ & Vertical Eddy Diffusion Coefficient & $\mathcal{U}(8,11)$ & $9.63^{+0.89}_{-0.98}$\\
    $x_{\text{R}_{\text{p}}}$ (unitless) & Radius Adjustment Factor & $\mathcal{U}(-0.15,0.15)$ & $-0.01^{+0.01}_{-0.01}$\\
    $\delta_{\text{NRS1}}/\text{ppm}$ & NIRSpec NRS1 Offset & $\mathcal{U}(-400,400)$ & $-107^{+8}_{-8}$ \\
    $\delta_{\text{NRS2}}/\text{ppm}$ & NIRSpec NRS2 Offset & $\mathcal{U}(-400,400)$ & $-68^{+7}_{-7}$\\
    $\log \text{P}_{\text{ref}}~\text{(bar)}$ & Reference Pressure & $\mathcal{U}(-6, 0.9)$ & $-3.24^{+1.98}_{-1.82}$\\
    \enddata
\end{deluxetable*}

We perform a Bayesian grid fitting analysis using \picaso on our model grid \cite[following the procedure presented in e.g.,][]{JWSTCO2,Barat25,Mukherjee25b}.
For the star and planet inputs, we use the median values calculated in Section \ref{sec:stellar}.
We interpolate over our model grid's chemical and thermal profiles, allowing each of the four input parameters to freely vary in our analysis.
We generate model spectra using the radiative transfer algorithm presented in \citet{toon1989rapid}.

During our Bayesian grid fitting analysis, we simulate clouds in KELT-7\,b's atmosphere using the open-source \texttt{VIRGA} cloud code \citep{Rooney22, Batalha26}.
Due to KELT-7\,b's atmospheric temperature, this is primarily corundum (Al$_2$O$_3$).
We first simulate cloud properties with the thermal properties of the clear grid model as inputs. 
Then, we generate model transmission spectra corresponding to these post-processed clouds\footnote{Because they are post-processed, these models do not account for the radiative effects of the clouds on the atmospheric thermal structure.}.
Our cloud model includes two parameters: the cloud sedimentation factor f$_{\text{sed}}$, defined as the ratio between the particle sedimentation velocity and the velocity of convective mixing, and the vertical eddy diffusion coefficient K$_{\text{zz}}$.
Lower values of f$_{\text{sed}}$ produce cloud structures with droplets lofted to high altitudes in the atmosphere due to less efficient sedimentation; similarly, higher values of K$_{\text{zz}}$ keeps larger cloud particles lofted in the atmosphere due to more increased vertical transport\citep{Rooney22,Ackerman01, Mukherjee25}.

In addition to our model- and cloud-related parameters, we also include as free parameters a radius adjustment factor $x_{\text{R$_{\text{p}}$}}$ such that the planet's radius R$_{\text{p}} = \text{R}_{\text{p},0}(1 + x_{\text{R$_p$}})$ \citep[e.g.,][]{Mukherjee25}, an offset between NIRISS/SOSS and each detector of NIRSpec/G395H, and the reference pressure P$_{\text{ref}}$; in total, our model has 10 parameters.
We describe the priors for our grid fit in Table \ref{tab:picaso}, and execute our Bayesian grid fitting analysis using \texttt{PyMultiNest} with 1000 live points.

\subsection{Equilibrium Retrieval and Grid Fit Results}
\begin{deluxetable*}{lcc}
\centering
\tablewidth{0pt}
\tablecaption{Equilibrium Retrieval Results}
\tablehead{
Parameter & \colhead{\poseidon} & \colhead{\pRT}
}
\startdata
    \hline
    Reference Radius ($\text{R}_{\mathrm{p, \, ref}}$/$\text{R}_{\text{Jup}}$) & $1.51^{+0.0005}_{-0.0005}$ & $1.55^{+0.001}_{-0.001}$\\
    \hline
    Atmospheric Temperature ($T_{\mathrm{ref}}$/K) & $2041^{+50}_{-33}$ & $1572^{+62}_{-68}$ \\
    $T(P)$ Profile Curvature 1 ($\alpha_{1}$/K$^{-\frac{1}{2}}$) & $1.00^{+0.65}_{-0.54}$ & ---  \\
    $T(P)$ Profile Curvature 2 ($\alpha_{2}$/K$^{-\frac{1}{2}}$) & $1.16^{+0.55}_{-0.54}$ & ---  \\
    $T(P)$ Profile Region 1 ($\log_{10} (P_{1}$ / bar)) & $-2.92^{+1.69}_{-1.76}$ & --- \\
    $T(P)$ Profile Region 2 ($\log_{10} (P_{2}$ / bar)) & $-3.27^{+1.86}_{-1.77}$ & --- \\
    $T(P)$ Profile Region 3 ($\log_{10} (P_{3}$ / bar)) & $0.23^{+1.15}_{-1.24}$ & --- \\
    Infrared Opacity ($\log_{10} (\kappa_{\text{IR}}$ / cm$^2$ g$^{-1})$) & --- & $-0.56^{+0.16}_{-0.17}$ \\
    Optical/Infrared Opacity Ratio ($\log_{10} (\gamma)$) & --- & $1.29^{+0.11}_{-0.11}$ \\
    \hline
    Cloud Top Pressure ($\log_{\rm{10}} (P_{\rm{cloud}}$ / bar)) & $-1.86^{+1.86}_{-2.38}$ & $-0.62^{+1.76}_{-0.45}$ \\
    Aerosol Volume Mixing Ratio ($\log_{\rm{10}} \text{Al}_2\text{O}_3$) & $-17.91^{+9.98}_{-7.96}$ & --- \\
    Aerosol Particle Size ($\log_{\rm{10}} (r_{\text{Al}_2\text{O}_3}$/$\mu$m)) & $-2.03^{+0.65}_{-0.64}$ & --- \\
    Slab Pressure Extent ($\Delta\log_{10} P_{\text{slab}}$) & $3.54^{+2.33}_{-2.29}$ & --- \\
    \hline
    Metallicity ($\log\text{M/H}$) & $1.91^{+0.06}_{-0.06}$ & $2.06^{+0.06}_{-0.06}$ \\
    C/O & $0.21^{+0.01}_{-0.01}$ & $0.73^{+0.01}_{-0.01}$ \\
    \hline
    Data Offset 1 ($\delta_{\mathrm{NRS1}}$/ppm) & $101^{+6}_{-6}$ & $89^{+8}_{-8}$  \\
    Data Offset 2 ($\delta_{\mathrm{NRS2}}$/ppm) & $59^{+6}_{-6}$ & $18^{+9}_{-8}$  \\
    \hline
\enddata
\tablecomments{$T_{\mathrm{ref}}$ refers to the top-of-atmosphere temperature for \poseidon, while for \pRT it represents the equilibrium temperature. Offsets are relative to NIRISS/SOSS.}
\label{tab:eq_results}
\end{deluxetable*}

\begin{figure*}
    \centering
    \includegraphics[width=\linewidth]{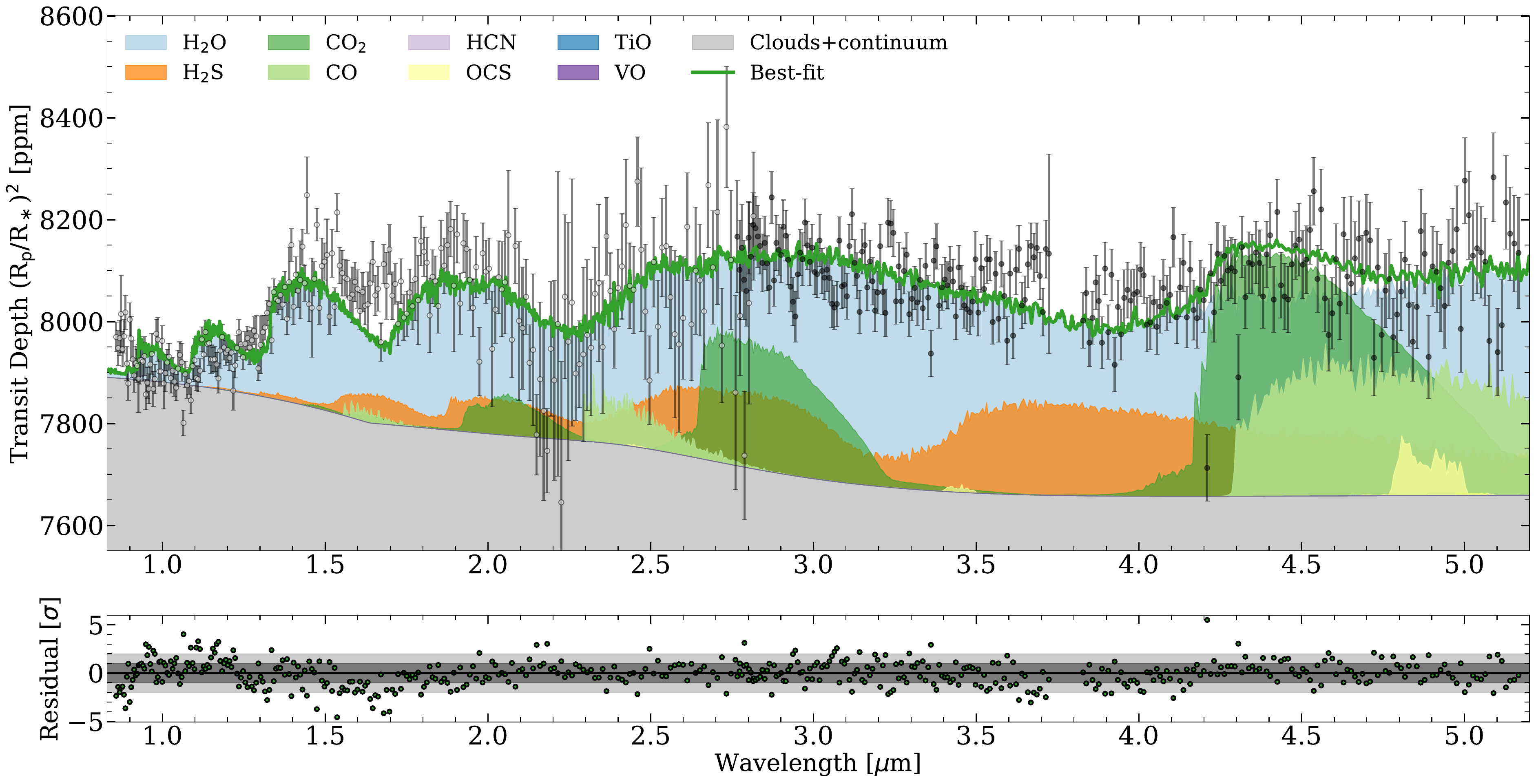}
    \caption{Summary of the results from our \picaso equilibrium chemistry Bayesian grid fitting analysis. In the upper panel we show a comparison between the retrieved transmission spectrum and the data used in our analysis. 
    We plot as the green line the best-fit retrieved spectrum and as black points the JWST NIRISS (unfilled circles) and NIRSpec (filled circles) data points.
    We overplot as shaded regions the spectral contribution for the best-fit spectrum.
    We apply separate offsets between NIRISS/SOSS and each detector of the archival JWST NIRSpec/G395H \eureka reduction presented in \citet{Ahrer25}.
    In the lower panel we show the residual between the best-fit model and observed transmission spectrum points as black points with green centers.
    We also plot as dark and light shaded regions the 1-$\sigma$ and 2-$\sigma$ difference regions, with the horizontal black line representing exact agreement.
    Our self-consistent equilibrium chemistry model grid retrieves a high atmospheric metallicity and low C/O ratio for KELT-7\,b.
    However, the best-fit model still fits the short-wavelength data poorly, relying on clouds to slightly mute features.}
    \label{fig:picaso_eq}
\end{figure*}

\begin{figure*}
    \centering
    \includegraphics[width=\linewidth]{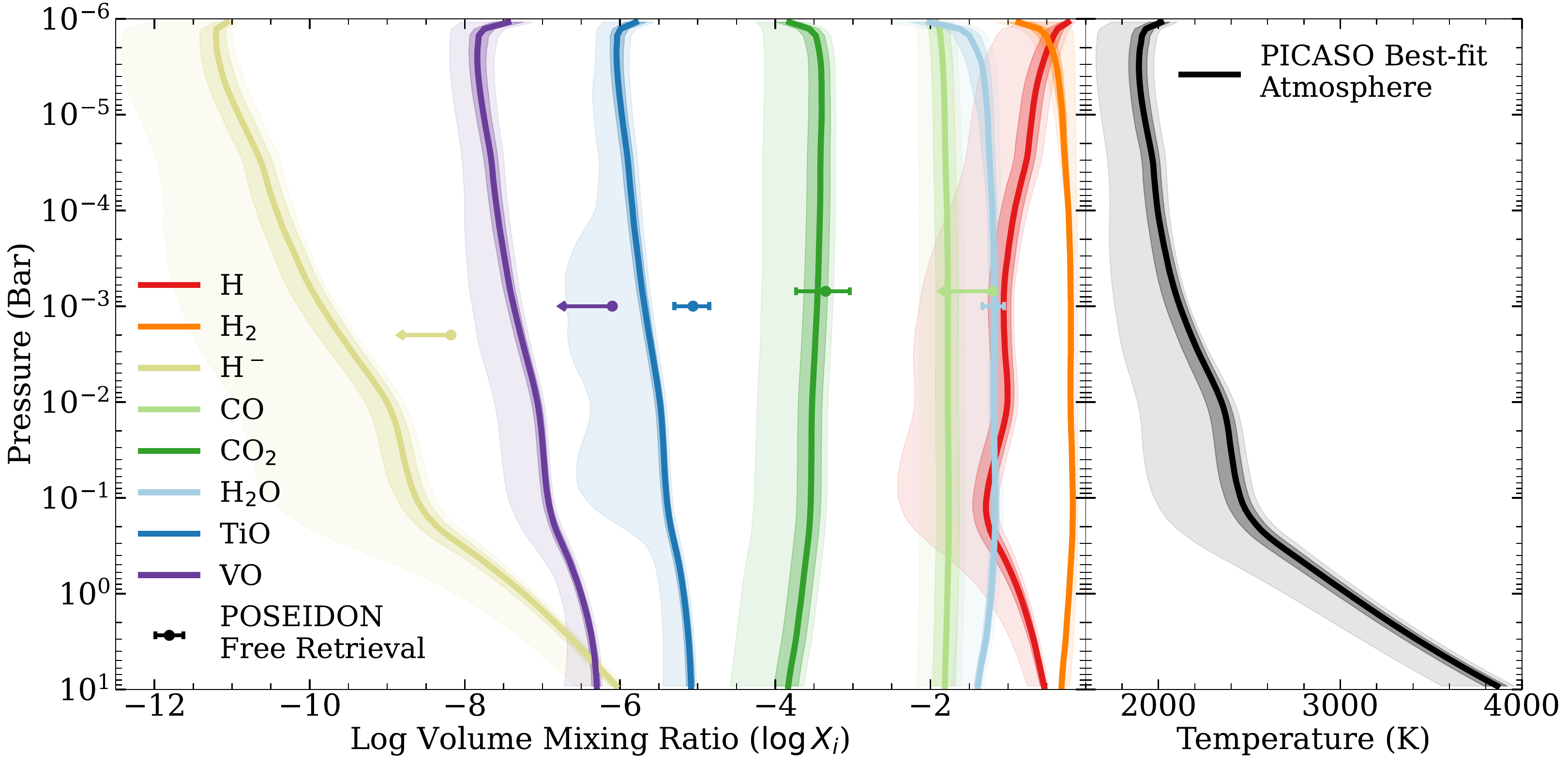}
    \caption{Comparison between our \poseidon free retrieval results and the best-fit \picaso equilibrium chemistry model. We plot as the solid lines several select chemical profiles (colored lines, left panel) and the $T(P)$ profile (black line, right panel) from our \picaso grid fit analysis. We show the 1-$\sigma$ and 2-$\sigma$ confidence intervals for these profiles as shaded polygons. We overplot as the error bars the volume mixing ratios from our \poseidon panchromatic free retrieval analysis, with left-pointing arrows corresponding to 2-$\sigma$ upper limits. Our free retrieval results are well in agreement with the physically self-consistent \picaso best-fit model, pointing to a metal-rich atmosphere near chemical equilibrium for KELT-7\,b.}
    \label{fig:chemcomp}
\end{figure*}

Each of our equilibrium chemistry retrievals and Bayesian grid fitting analysis on KELT-7\,b's transmission spectrum prefers a high atmospheric metallicity.
We list the fitted values for each code in Table \ref{tab:picaso} (for \picaso) or in Table \ref{tab:eq_results} (for \poseidon and \pRT).
Weighting each posterior equally to agglomerate our results\footnote{We do this as we cannot calculate Bayes factors between the posteriors, and we do not prefer one retrieval over another.}, we find that between our three codes we infer a metallicity of $\log \text{M/H}=1.96^{+0.13}_{-0.10}$, corresponding to $\text{M/H}=92^{+23}_{-22}~\times$~Solar. 
We show the posterior corner plots for these analyses in Appendix \ref{app:eqcorner}, as well as a spectral contribution for our \picaso analysis in Figure \ref{fig:picaso_eq}.
As we show in Figure \ref{fig:chemcomp}, the VMRs (as well as the 2-$\sigma$ upper limits) inferred by our free retrieval analysis are also consistent with the chemical profiles from our \picaso analysis.

Despite the similarity in the inferred atmospheric metallicity, our three analyses differ in their inferred C/O ratios: while \poseidon prefers a low value at the edge of its prior range, \pRT prefers a more moderate value while \picaso allows a broad, uninformative range.
Similar to the results of \citet{Ahrer25}, we rule out high C/O values above C/O~$\sim0.9$.
The higher C/O inferred by \pRT may in part be due to its exclusion of several oxygen-containing species that \poseidon and \picaso include, namely OCS, SO$_2$, TiO, VO, and SiO.
The different inferred C/O ratios could also be related to how each code calculates C/O in reference to the metallicity: while \pRT sets metallicity from C/H and varies oxygen based on the C/O ratio, \poseidon instead sets metallicity from O/H and varies carbon.
Meanwhile, \picaso uses the total amount of C and O to set the metallicity, varying C/O while keeping their total number constant.

Our \pRT equilibrium retrieval also differs from our \poseidon retrieval in that its inferred atmospheric equilibrium temperature is $\sim500$ K cooler than KELT-7\,b's zero-albedo equilibrium temperature.
On the other hand, our \poseidon retrieval finds an atmospheric temperature much closer to the zero-albedo value.
This could further explain the discrepancy between the C/O ratios inferred by \poseidon and \pRT as equilibrium chemistry would require relatively more carbon at lower temperatures to produce the same 4--5 $\mu$m CO and CO$_2$ features that a hotter atmosphere would.
Though our \picaso analysis does not provide any constraints on the intrinsic temperature, it does prefer high values of the heat redistribution parameter f$_{\text{rfacv}}$, suggesting efficient heat redistribution.
Like our \poseidon and \pRT retrievals, our \picaso analysis also does not constrain the presence of clouds, with wide posterior ranges for the cloud parameters f$_{\text{sed}}$ and K$_{\text{zz}}$.

\begin{figure*}[ht!]
    \centering
    \includegraphics[width=\linewidth]{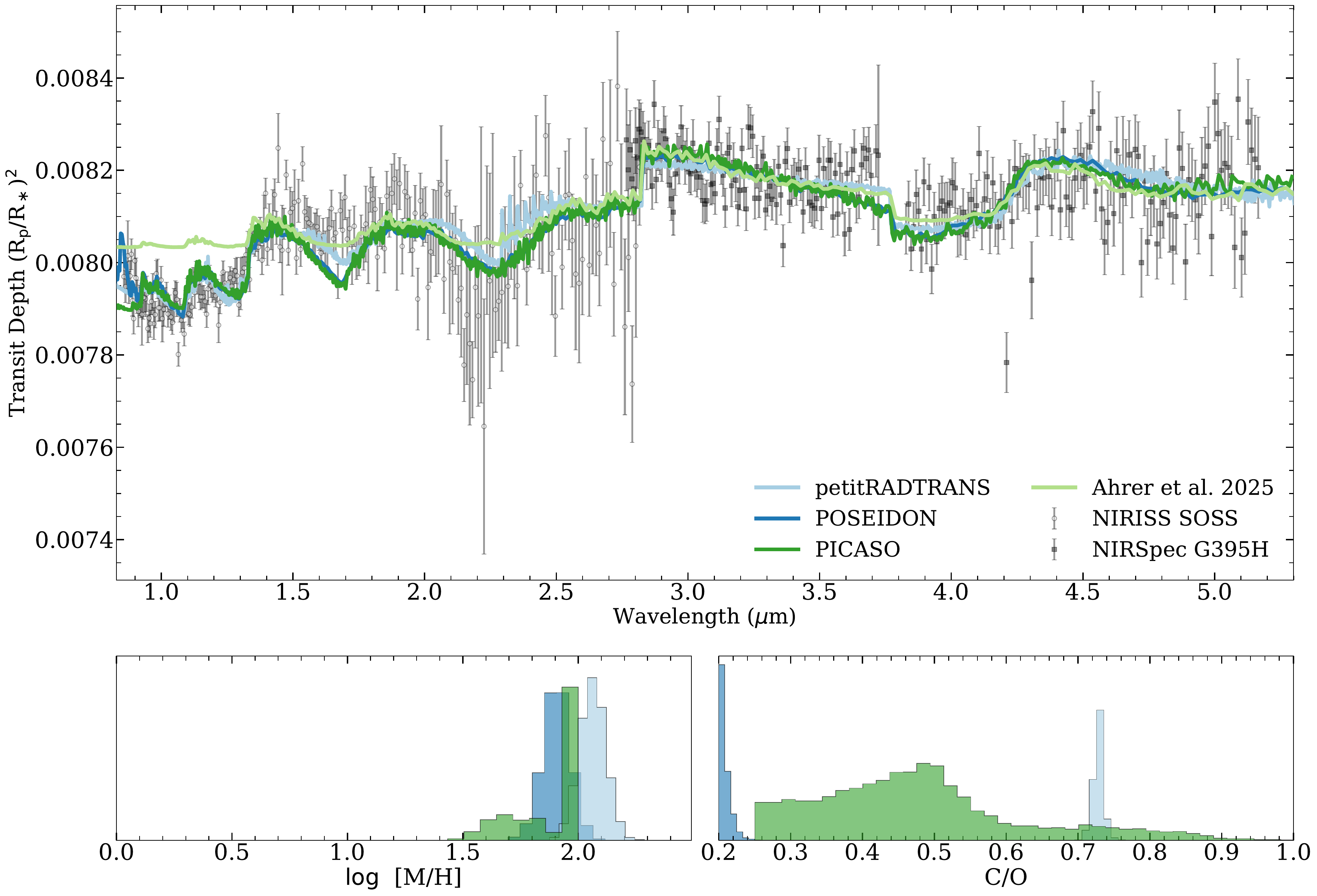}
    \caption{Comparison between equilibrium retrievals performed using \pRT (light blue) and \poseidon (dark blue), as well as our \picaso Bayesian grid fitting analysis (dark green), on the JWST NIRISS and NIRSpec transmission spectra of KELT-7\,b. 
    In the upper panel we plot as unfilled points the NIRISS transmission spectrum presented in this work and as filled points the NIRSpec/G395H transmission spectrum presented in \citet{Ahrer25}. 
    We overplot as solid lines the best-fit (for \pRT and \picaso) or median (for \poseidon) spectra.
    The jumps present in each spectrum show the retrieved offsets from each code's results posterior; note that they differ for each. 
    We additionally overplot as the light green line the median spectrum from \citet{Ahrer25}'s \poseidon equilibrium chemistry retrieval on the \eureka $R\sim400$ NIRSpec/G395H transmission spectrum.
    We apply an offset to this spectrum to better match the NIRISS data, but its high cloud deck causes it to not fit the short-wavelength data.
    In the lower panels we show posteriors for atmospheric metallicity (left) and C/O ratio (right) from each analysis presented in this work. 
    While all three codes prefer a metal-rich atmosphere, \pRT prefers a moderate C/O ratio of $0.73^{+0.01}_{-0.01}$ whereas \poseidon approaches the lower end of its prior distribution and \picaso provides no strong constraint; combined together, we rule out C/O $\gtrsim 0.9$.}
    \label{fig:retrievalcomparison}
\end{figure*}

The overall shapes of our best-fit (\pRT and \picaso) and median (\poseidon) spectra, which we compare in Figure \ref{fig:retrievalcomparison}, are quite similar to each other.
They differ slightly in their best-fit offsets between NIRISS and the two detectors of NIRSpec as well as the shapes of individual features, but this is in part due to different choices of opacity tables or line lists.
Notably, each of our analyses fits the data rather poorly at shorter wavelengths, preferring H$_2$O features when the actual data appear more muted.

\section{3D Climate Modeling}\label{sec:gcm}
\begin{deluxetable}{lcc}
\label{tab:input_parameter_UM}
\centering
\tablewidth{0pt}
\tablecaption{Stellar and planetary parameters for GCM input}
\tablehead{
Parameter & Value 
}
\startdata
Stellar irradiance [W\,m$^{-2}$] & 3.509$\times$10$^{6}$ \\
Stellar constant at 1\,AU [W\,m$^{-2}$] & $7.587 \times 10^3$\\
Inner radius [m] &  1.115$\times$10$^{8}$\\
Domain height [m] & 9$\times$10$^{6}$\\
Semi-major axis [AU] & 4.65$\times$10$^{-2}$ \\
Orbital period [Earth day]  & 2.73 \\
Rotation rate [rad\,s$^{-1}$] &  2.659$\times$10$^{-5}$ \\
Surface gravity [m\,s$^{-2}$] & 12.4 \\
Specific gas constant [J\,K$^{-1}$\,kg$^{-1}$] & $3.196 \times 10^3$ \\
Specific heat capacity [J\,K$^{-1}$\,kg$^{-1}$] & $1.2356 \times 10^4$ \\
Metallicity & 10$\times$ Solar\\
C/O ratio & 0.55\\
\enddata
\end{deluxetable}

Our \texttt{catwoman} two-limb spectra of KELT-7\,b did not yield significant deviations from the spherical transmission spectrum.
To assess a physical scenario for which this would be possible, we perform 3D General Circulation Model (GCM) climate simulations of the atmosphere of KELT-7\,b using the Met Office \texttt{Unified Model} (UM). 
The UM consists of the dynamical core, Even Newer Dynamics for General atmospheric modeling of the environment (ENDGame), and solves the non-hydrostatic, full deep-atmosphere equations of motion with varying gravity within the atmosphere \citep[see][for discussion]{Wood14,Mayne14a,Mayne14b,Mayne17}. 
The UM has been adapted to study a range of gas giant exoplanets \citep{Christie21,Zamyatina23,Zamyatina24,Mak25}. 
The UM is also coupled with the 2-stream radiative transfer scheme from the ``Suite of Community RAdiative Transfer codES'' \citep[\texttt{Socrates}, based on][]{Edwards_and_Slingo_1996} and solves for the gaseous absorption from \ce{H2O}, \ce{CH4}, \ce{CO}, \ce{CO2}, \ce{HCN}, \ce{NH3}, \ce{Li}, \ce{Na}, \ce{K}, \ce{Rb}, and \ce{Cs}, as well as collision-induced absorption from \ce{H2}-\ce{H2} and \ce{H2}-\ce{He} from and Rayleigh scattering due to \ce{H2} and \ce{He} using the correlated-\textit{k} method \citep[see details of line list opacity sources in Appendix \ref{app:opacities} in][]{Zamyatina23}. 
We do not include H$^-$ opacity as our free retrieval analysis disfavors its presence.

We initialize the UM with a day side mean $T(P)$ profile from the 1D radiative-convective-chemistry code ATMO \citep[see for example][]{Tremblin15, Drummond16,Drummond19} assuming thermochemical equilibrium for the species in the C/H/N/O network of \citet{Venot12}. 
For our model stellar electromagnetic spectrum, we use a BT-Settl model \citep{Allard2014}.
To correct for thermal inversions due to initial conditions \citep[see][]{Drummond20,Zamyatina24} caused by the slow evolution of the deep atmosphere in GCMs \citep{Sainsbury-Martinez19}, we alter the $T(P)$ profiles from ATMO in the following ways: (1) by increasing the $T(P)$ structure by 500\,K, (2) by applying an adiabatic correction between pressures of 1--10\,bar, and (3) by smoothing the adiabatic region to an isothermal temperature of 3500\,K, between 10\,bar and the bottom of the model domain. 
We subsequently use the chemical kinetics and equilibrium schemes to study the climate and chemical species evolution of the planet. 
The kinetics chemistry scheme uses the C/H/N/O chemical network from \citet{Venot19}, which consists of 30 species and 181 reversible thermo-reactions (excluding photochemistry) and solves for the stiff ordinary differential equations which calculate the production and loss of chemical species \citep[see][for further details]{Drummond20,Zamyatina23,Zamyatina24}. 
The chemical equilibrium scheme uses the Gibbs minimization scheme to calculate the same chemical species abundance listed in the chemical kinetics scheme, but within the local grid cell assuming chemical equilibrium \citep[see][for details]{Drummond18a}.

\begin{figure*}
    \centering
    \includegraphics[width=\linewidth]{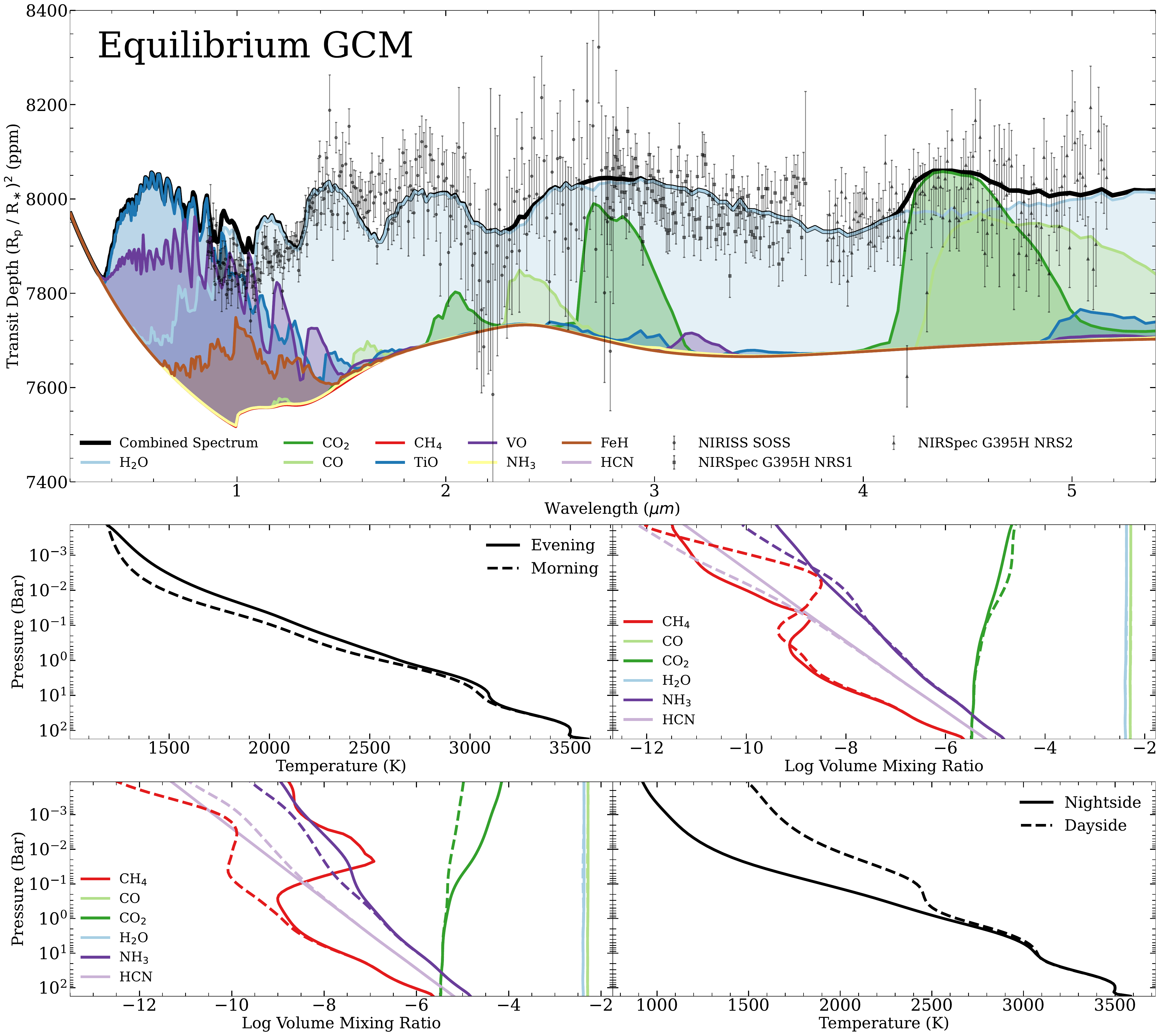}
    \caption{Results of our aerosol-free equilibrium chemistry general circulation model of KELT-7\,b's atmosphere. In the top panel we show the limb-averaged spectral contribution plot of our model with the JWST NIRISS and NIRSpec data overlaid for comparison purposes. We apply offsets to each dataset to best match the averaged model transmission spectrum. In the lower panels we show the thermal (two lines) and chemical (several lines) profiles for the evening and morning limbs (middle) as well as the day side and night side (bottom). Due to a strong equatorial super-rotating jet, our model's $T(P)$ profile is similar between the limbs.}
    \label{fig:eq_gcm}
\end{figure*}

\begin{figure*}
    \centering
    \includegraphics[width=\linewidth]{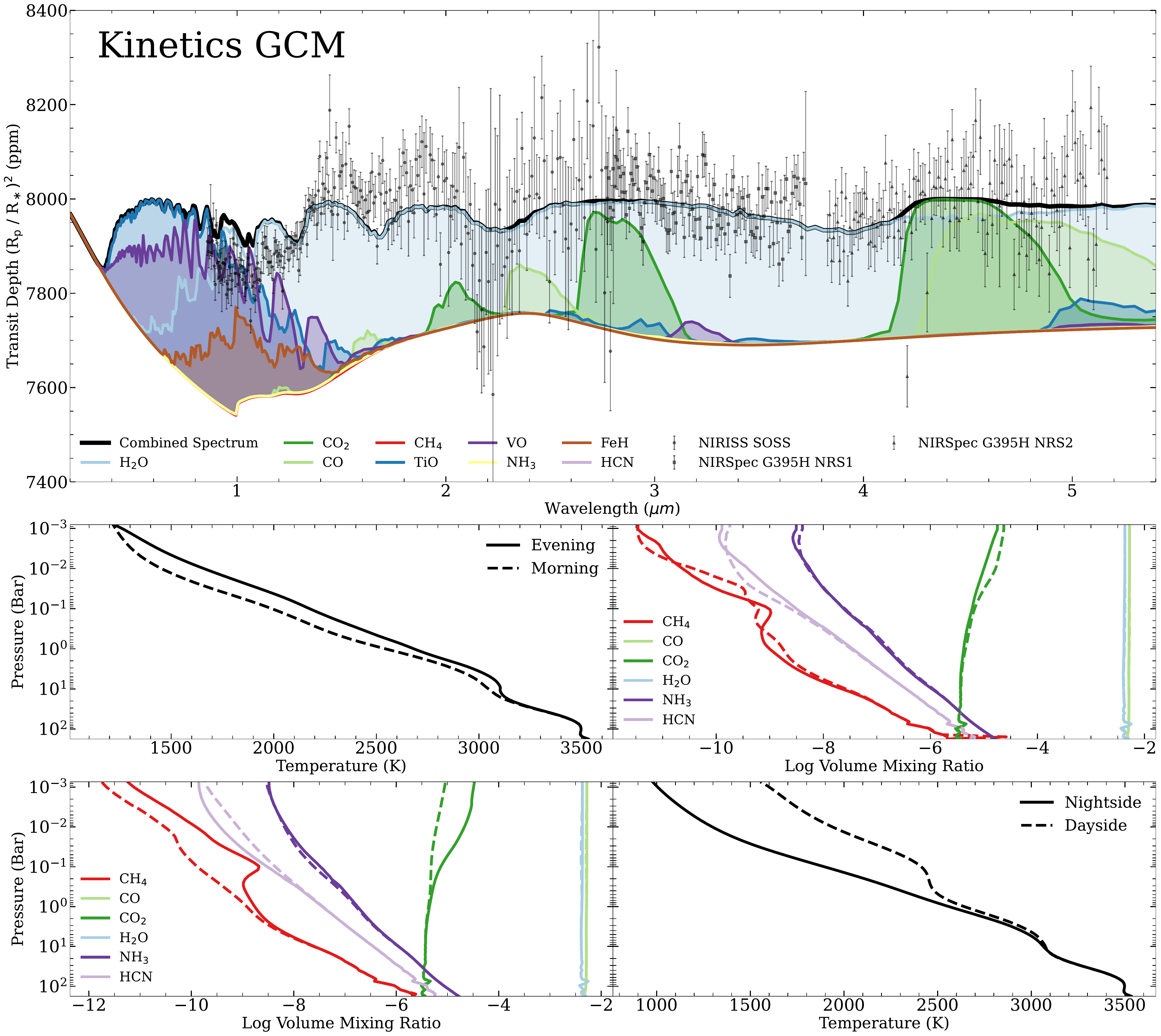}
    \caption{Results of our aerosol-free chemical kinetics general circulation model of KELT-7\,b's atmosphere. The format is the same as Figure \ref{fig:eq_gcm}. Similar to the equilibrium chemistry setup, our model's chemistry and $T(P)$ profile are similar between the limbs due to the strong horizontal jet homogenizing material distribution, further explaining the lack of limb asymmetry seen during our NIRISS/SOSS reduction.}
    \label{fig:gcm}
\end{figure*}

\begin{figure}
    \centering
    \includegraphics[width=\linewidth]{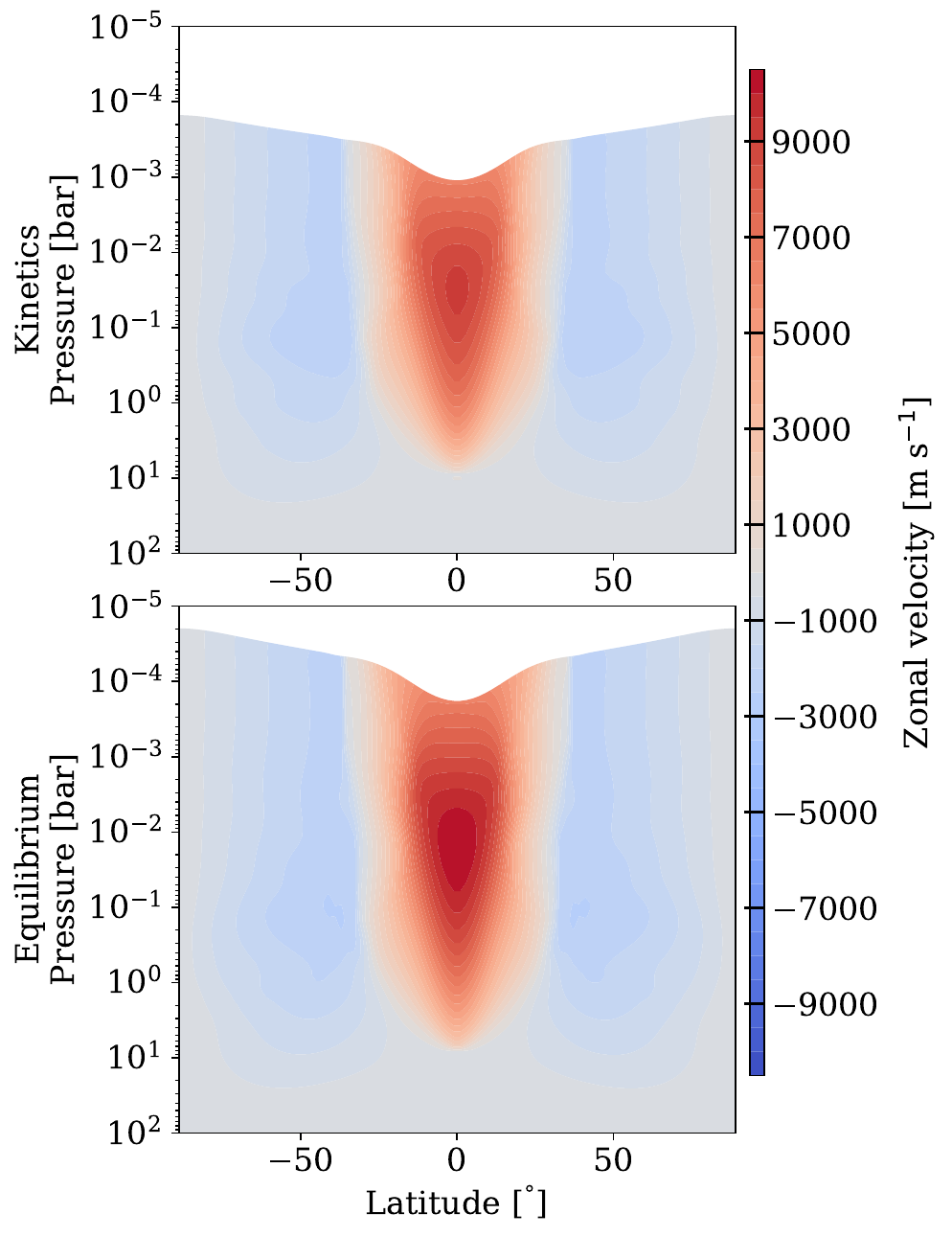}
    \caption{Zonally-averaged flow of our aerosol-free chemical kinetics and equilibrium chemistry setup within the general circulation model of KELT-7\,b's atmosphere. The strong wind speed facilitates efficient horizontal mixing and homogenies the heat and material distribution between the two planetary limbs.}
    \label{fig:GCM_zonal_wind}
\end{figure}

We list the stellar and planetary input parameters we use for the GCM in Table~\ref{tab:input_parameter_UM}. 
During the spin-up phase of the kinetic chemistry setup, we vary both the radiative and dynamical time steps between 20 and 30\,s before increasing to 60 s once the simulation becomes more stable.
For the equilibrium chemistry setup, we opt for the radiative and dynamical time steps to be 30 and 20\,s, respectively, throughout the entire simulation runtime.
To maintain numerical stability, we follow the approach from \citet{Zamyatina23} and adopt a filtering constant $K$ of $\sim$0.055 to help conserve axial angular moment, weaken the longitudinal filtering that is applied to the horizontal wind and prevent the formation of a retrograde equatorial flow.
We also adopt the depth of the sponge layer to be 0.75 relative to the model domain height, and the vertical damping coefficient to be 0.25\footnote{For detailed discussions on the longitudinal filtering and sponge layer treatment, see \citet{Christie24}.}.
All simulations are cloud- and haze-free and are performed using a horizontal grid spacing of 2.5$^\circ$ in longitude and 2$^\circ$ in latitude, with a vertical grid of 80 equally-spaced model levels.
To further maintain numerical stability, the lowest pressure of our simulation can only reach $\sim10^{-3}$ on the day side and $\sim10^{-4}$\,bar on the night side for the kinetics chemistry setup, and $\sim10^{-4}$ on the day side and $\sim3\times10^{-5}$\,bar on the night side for the equilibrium chemistry setup. 

We run the simulation for at least 800 Earth days to reach a pseudo-equilibrium, diagnosed by the plateauing of changes in the globally averaged top-of-atmosphere net energy flux, the maximum vertical and horizontal wind speeds, and the abundances of \ce{CH4}, \ce{CO}, \ce{H2O}, \ce{CO2}, \ce{NH3} and \ce{HCN} relative to their initial state. 
We temporally-average the results from the last 50\,Earth days and present them within this study.
We further calculate the transmission spectrum with \poseidon, which we show in Figure \ref{fig:eq_gcm} for equilibrium and Figure \ref{fig:gcm} for kinetics.
We use the GCM output $T(P)$ profile as well as the species abundance profiles of \ce{H2O}, \ce{CO2}, \ce{CO}, \ce{CH4}, \ce{NH3}, and \ce{HCN} as input files, whereas the abundance of \ce{TiO}, \ce{VO} and \ce{FeH} are calculated using the \poseidon package's \texttt{FASTCHEM} grid \citep{FASTCHEM} with the GCM output $T(P)$ profile.

Both the kinetic and equilibrium chemistry simulations show that the atmosphere of KELT-7\,b exhibits a very strong equatorial super-rotating jet, as shown in Figure \ref{fig:GCM_zonal_wind}, a circulation pattern commonly found among tidally-locked planets due to the strong day-night temperature-pressure gradient.
The maximum wind speed in our simulations reach $\sim9213$\,m\,s$^{-1}$ and $\sim10498$\,m\,s$^{-1}$ for the kinetics and equilibrium chemistry setup, respectively. 
Such vigorous atmospheric circulation efficiently redistributes heat across longitudes through enhanced horizontal mixing.
We note that the wind speeds obtained from our GCM simulations are not quantitatively predictive and other models may produce different results. 
Moreover, the presence of the super-rotating jet does not necessarily imply negligible temperature differences between the two planetary limbs. 
Previous GCMs have demonstrated that highly irradiated hot-Jupiters could exhibit a large limb asymmetries due to a less efficient heat redistribution, based on the balance between the radiative, advective and wave propagation timescales \citep{Roth24}. 
In our simulations here, this balance produces a similar thermal structures between the two planetary limbs and give rise to a similar chemical abundance profiles, as demonstrated by the equilibrium chemistry setup in Figure \ref{fig:eq_gcm}. 
The efficient horizontal transport associated with the jet further reduces abundance contrasts between species as demonstrated in the kinetic chemistry setup in Figure \ref{fig:gcm}.
This combined effect results in two-limb spectra that resemble the spherical transmission spectrum.

\section{Discussion} \label{sec:disc}
We have performed a full analysis of KELT-7\,b's panchromatic transmission spectrum using a new method to correct for charge migration in the NIRISS data. 
Our free retrieval analysis detects H$_2$O, CO$_2$, and the set of 6 high-temperature species (dominated by TiO) as well as unocculted starspots.
Notably, we do not detect H$^-$ opacity or a cloud deck.
Meanwhile, our equilibrium chemistry analyses together prefer a super-stellar $\text{M/H}=92^{+23}_{-22}~\times$~Solar ($\sim32\times$ Stellar), but poorly constrain the C/O ratio to C/O~$\leq0.9$.

Our retrieval results on the panchromatic transmission spectrum of KELT-7\,b differ strongly from previous results on subsets of our data.
For example, \citet{Gascon25}, which analyzed the HST data as well as photometric points from Spitzer, detected H$^-$ opacity seven orders of magnitude more abundant than predicted by equilibrium chemistry.
While we do not strongly detect H$^-$ opacity in our free retrieval, the retrieved VMR's 1-$\sigma$ range, $\log X_{\text{H$^-$}} = -10.11^{+1.20}_{-1.16}$, is within an order of magnitude of the equilibrium value.
This places H$^-$, and by extension, KELT-7\,b's atmosphere, much more in line with the predictions from equilibrium chemistry.
Similarly, \citet{Ahrer25}'s analysis on NIRSpec and the HST data found only tentative evidence for CO$_2$ and H$_2$O; they also were unable to constrain the presence of any high-temperature species.
On the other hand, our free retrieval finds strong detections of CO$_2$,  H$_2$O, and the high-temperature species.
We attribute this to the strong constraint on H$_2$O from NIRISS setting the spectrum's baseline deeper than was observed in NIRSpec.
This implies that the low-amplitude features in NIRspec are dominated by molecular opacity rather than a cloud deck.
In the interpretation of their equilibrium retrievals, \citet{Ahrer25} outlined the possibility that either a high-altitude cloud deck or a low-metallicity atmosphere could best explain their NIRSpec/G395H transmission spectrum. 
Our equilibrium analyses on the combination of NIRISS and NIRSpec data again contrasts with this, preferring no clouds and a high-metallicity atmosphere; as we show in Figure \ref{fig:retrievalcomparison}, the median \poseidon model for the $R\sim400$ \eureka NIRSpec/G395H data presented in \citet{Ahrer25} is too cloudy to fit the short-wavelength NIRISS data.
Given the significant differences between this previous work and ours, we emphasize that broad wavelength coverage beyond a single JWST instrument is crucial for the comprehensive study of exoplanet atmospheres.

\begin{figure}
    \centering
    \includegraphics[width=\linewidth]{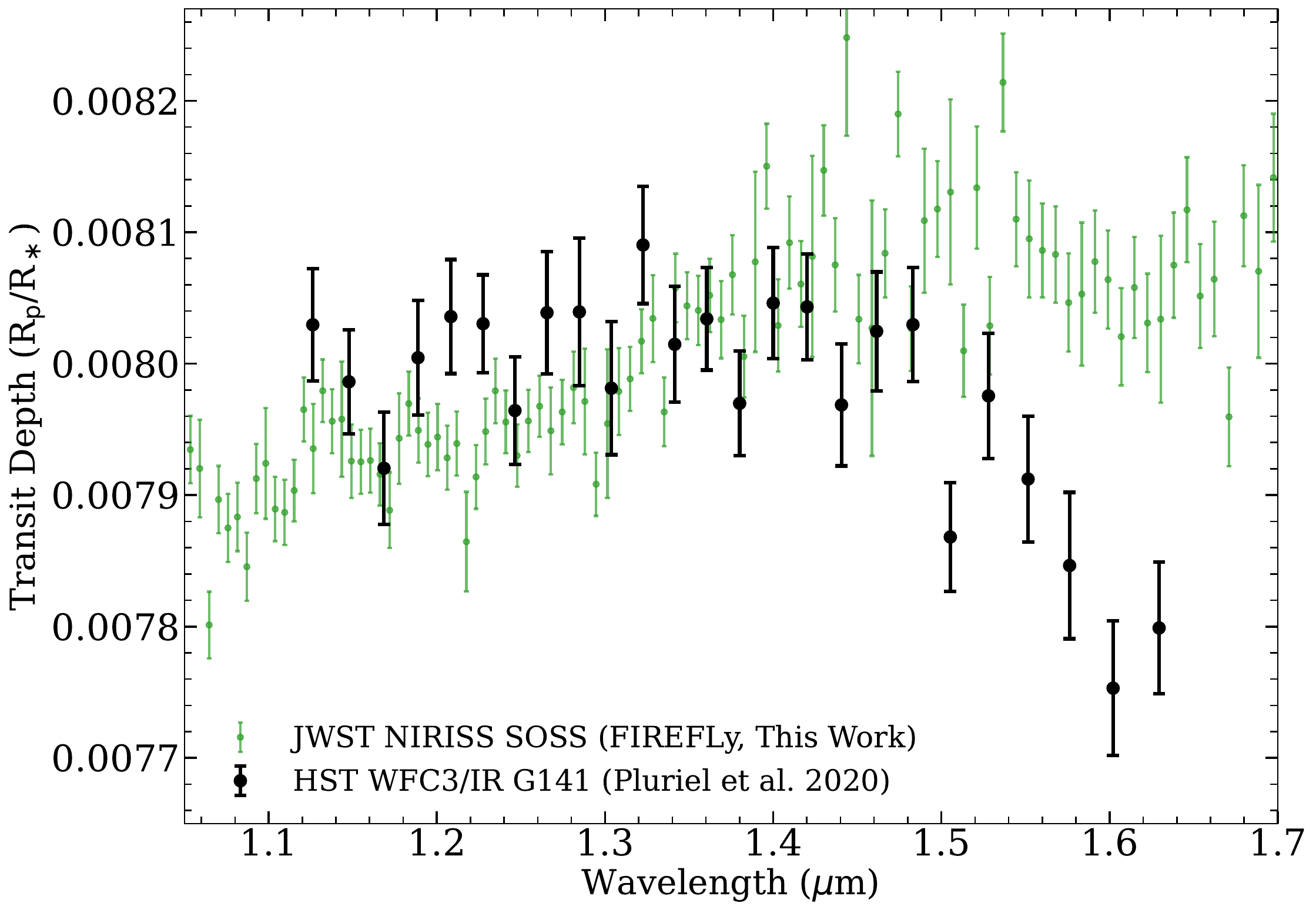}
    \caption{Comparison between our JWST NIRISS transmission spectrum and the HST WFC3/IR G141 transmission spectrum presented in \citet{Pluriel20}. We plot as green points our \firefly NIRISS/SOSS reduction and as black points the WFC3/IR G141 data with an offset applied. While the data are broadly consistent between 1.1--1.5 $\mu$m, the WFC3/IR G141 data show shallower transit depths relative to NIRISS beyond 1.5 $\mu$m. As many other hot Jupiters' G141 transmission spectra also exhibit a similar feature at these wavelengths, the shallower transit depths may be caused by an unaccounted-for instrumental systematic.}
    \label{fig:g141_comparison}
\end{figure}

There is a distinct difference between the shape of our NIRISS transmission spectrum and the HST WFC3/IR G141 transmission spectrum of KELT-7\,b between 1.4--1.7 $\mu$m. 
We show this difference between our transmission spectrum and \citet{Pluriel20}'s transmission spectrum in Figure \ref{fig:g141_comparison}.
This feature was previously used to claim H$^-$ opacity in the planet's atmosphere, but more precise follow-up from JWST disfavors this claim. 
We note that the JWST data longward of 1.5 $\mu$m, which shows the largest disagreement, is not affected by charge migration.
This could indicate that an unaccounted-for instrumental systematic may be present among some WFC3/IR G141 observations.
Indeed, many WFC3/IR G141 observations of hot Jupiters exhibit a similar shape as KELT-7\,b's at these wavelengths (e.g., \citet{Edwards23} Figure 2, lower right, and \citet{Lothringer25}).
A future reanalysis effort should therefore be undertaken to correct for this, or to otherwise observe potentially affected planets' spectra with NIRISS to quantify the extent of the issue.

The metallicity we infer for KELT-7\,b is on the higher end of the plausible range for planets of KELT-7\,b's mass and KELT-7's metallicity \citep{Chachan25} \citep[Though the higher end of the mass distribution does show more variance in metallicity; e.g.,][]{Lothringer26}.
This high metallicity is not caused by clouds muting features outside of the morning/evening terminators; when we combine a flat line with a lower metallicity atmosphere's transmission spectrum to simulate the effects of patchy clouds, we are still unable to fit the 1--1.5 $\mu$m region.
As KELT-7\,b's radius is inflated, reconciling a high metallicity with radius inflation could require a reassessment of the relationship between atmospheric temperature and efficiency of heating for UHJs \citep[e.g.,][]{Thorngren18}.
Alternatively, additional heating mechanisms \citep[especially ones that cause shallow heating;][]{Schmidt26b} or enhanced atmospheric opacity from the high metallicity \citep[e.g.,][]{Burrows2003, Burrows07} may help explain KELT-7\,b's peculiar situation.
Emission spectra or phase curve observations of KELT-7\,b would be able to confirm the metallicity we have inferred and probe the thermal structure of KELT-7\,b to assess its degree of peculiarity.

It is difficult to contextualize the atmospheric metallicity we infer for KELT-7\,b due to physical limitations imposed by the host star KELT-7.
As the host star's photospheric abundances are the most accurate proxy for protoplanetary disk abundances, detailed analysis of the host star's carbon and oxygen abundances is necessary to obtain an accurate inference of formation location \citep[e.g.,][]{Reggiani22,Reggiani24}.
However, our stellar analysis of KELT-7 is limited by the degree of rotational broadening present in its ground-based high-resolution optical spectrum.
This precludes us from inferring KELT-7\,b's C/O ratio relative to its host star.
Given our high atmospheric metallicity inference, it is still likely that KELT-7\,b's metallicity is super-stellar, but we are unable to quantify the degree to which this is the case.

We must appeal to formation in a high-metallicity environment or late-stage metal enhancement of KELT-7\,b's atmosphere to explain its super-stellar atmospheric metallicity in a planet formation context.
While this could imply that it formed near an ice line, this would also have resulted in a lower envelope metallicity and a super-solar C/O ratio unless it accreted planetesimals late in its formation \citep{Madhusudhan17}.
Alternatively, KELT-7\,b's formation location could have become metal-enriched due to pebble drift \citep{Booth17, Bitsch23} or photoevaporation in the protoplanetary disk \citep{Lienert24}.
Inefficient planetesimal accretion could also contribute to a high atmospheric metallicity for KELT-7\,b \citep{Danti23}.

Disk migration remains the most plausible formation channel for KELT-7\,b over in situ formation or high-eccentricity migration.
Though massive, metal-rich inner disks are necessary for in situ formation to occur \citep{Bodenheimer00, Batygin16}, the sublimation of solids in this part of the parent protoplanetary disk does not preferentially enrich the planet in metals.
In situ formation scenarios would therefore produce a planet with a metallicity on par with the metallicity of the host star.
While its aligned orbit \citep{Zhou16} points towards a dynamically cool formation pathway, it is still possible that KELT-7\,b underwent high--eccentricity migration as this formation channel does not always cause a high stellar spin-orbit misalignment \citep[e.g.,][]{Chatterjee08}.
Other system properties \citep[such as orbital period; e.g.,][]{Schmidt26a} could then be useful differentiators of formation location or used as priors in atmospheric inference studies to improve their results' significance.
While disk migration is possible, it is only likely to have occurred if the planet formed early in a massive, long-lived disk and the disk inner edge was slightly closer in than what is theoretically expected.
Disk migration is relatively inefficient for planets around higher mass stars ($>1$ M$_\odot$), and disk lifetimes are shorter \citep{Johnston24}.
KELT-7\,b's current semimajor axis, 0.04 AU, is also close or slightly interior to KELT-7's protoplanetary disk magnetospheric truncation radius of 0.05 AU \citep[$r_{in} \approx 0.06 (\frac{M_\ast}{2.4~M_{\odot}})^{1/3}$ AU;][]{Mulders15}. 
As KELT-7's rotation period is faster than KELT-7\,b's orbital period \citep{Tabernero22}, angular momentum is being transferred from the rotation of the star to the orbit of the planet.
If KELT-7\,b indeed formed via this pathway, then KELT-7's parent protoplanetary disk must have truncated closer to the star than theory would suggest.

Finally, our \poseidon free retrieval shows ``decisive'' evidence for stellar contamination from the shape of the HST WFC3/UVIS G280 transmission spectrum.
However, KELT-7 is a mature F dwarf above the Kraft break and therefore not expected to possess a large convective envelope.
It is therefore unlikely a priori for it to have strong magnetic activity and inhomogeneities that cover $\sim25$\% of its surface as inferred by our free retrieval.
Nevertheless, KELT-7's rotation period is well-constrained to be close to half of KELT-7\,b's orbital period from brightness modulations observed by the Transiting Exoplanet Survey Satellite (TESS) \citep{Tabernero22}, and is similarly constrained via the combination of its high spectroscopic v$\sin i$ and the inferred radius from our analysis in Section \ref{sec:stellar}.
If this inference of unocculted stellar inhomogeneities is in part physical, then they may be caused by magnetic interactions between KELT-7\,b and its host star.
Further ultraviolet observations will be necessary to determine whether the observed shallow transit depths are caused by this, as opposed to an uncorrected-for instrumental systematic.

\section{Conclusion} \label{sec:conclusion}
We performed a panchromatic transmission spectrum analysis of the ultra-hot Jupiter KELT-7\,b, incorporating data from HST WFC3/UVIS G280, JWST NIRISS/SOSS, and JWST NIRSpec/G395H.
After identifying nonlinearity in the NIRISS observations, we hypothesized that charge migration was to blame, and modified the standard \texttt{jwst} reduction pipeline to mitigate its effects.
Finding that the charge migration-affected region was recovered, we then re-reduced the HST WFC3/UVIS G280 observations to allow for a coarser binning scheme and performed both free and equilibrium retrieval analyses on KELT-7\,b's transmission spectrum.
In the process, we also found that the NIRISS transmission spectrum's shape differed from the HST WFC3/IR G141 spectrum. This may be an instrumental systematic that should be considered in future studies of those data.
The late-ramp-fit method we demonstrated here for NIRISS should be thoroughly tested with appropriate calibration data to ensure its fidelity and the accuracy of the transit depths calculated with it; it should also be tested with other JWST instruments to ensure its applicability for detectors with fewer pixels.
If these are confirmed, then it will be applicable to many JWST datasets near detector brightness limits, maximizing the science output of these observations.

Our free retrievals on the full 0.2--5.1 $\mu$m spectrum with \poseidon showed decisive evidence for H$_2$O, strong evidence for CO$_2$, decisive evidence for high-temperature species like TiO, as well as decisive evidence for unocculted faculae (though the inferred stellar contamination is likely unphysical).
However, we did not find evidence for H$^-$ opacity or clouds.
We used these results to inform an equilibrium retrieval analysis across three different codes, including a Bayesian grid fitting analysis on physically self-consistent atmosphere models with \picaso and forward model-based retrievals with \poseidon and \pRT.
Through these retrievals we constrained the atmospheric metallicity to a super-stellar $92^{+23}_{-22}~\times$~Solar, but with discrepant C/O ratios that we attribute to differences in the inferred atmospheric temperature and molecules considered.
These retrievals also did not constrain the presence of clouds in KELT-7\,b's atmosphere.
The high super-stellar metallicity may indicate that KELT-7\,b accreted metals late in its formation; on the other hand, the poorly-constrained C/O ratio offers little insight into KELT-7\,b's formation history.
After finding no evidence of limb asymmetry in our NIRISS reduction, we calculated several 3D climate models of KELT-7\,b to assess physical scenarios for which this would occur.
We found that a strong super-rotating jet can facilitate horizontal mixing, homogenizing the planet's terminators and transit in the process.

This comprehensive analysis of KELT-7\,b would not have been possible without the panchromatic spectra provided by HST and JWST.
Our work has resolved several outstanding questions posed by previous work on KELT-7\,b while also opening new ones about both the nature of the planet and the science now accessible with JWST and HST.
Future studies of ultra-hot Jupiters will benefit from broader wavelength coverage to also obtain a clear view of these extreme planets' atmospheres.

\section*{Acknowledgments}
We thank the anonymous referee for their helpful comments.
We also thank Drake Deming and Michael Radica for helpful comments about our procedure.
Stephen P. Schmidt is supported by the National Science Foundation Graduate Research Fellowship Program under Grant No. DGE2139757.
Mei Ting Mak acknowledges support from the Croucher Postdoctoral Fellowship, funded by the Croucher Foundation.
Nathan J. Mayne acknowledges support from a UKRI Future Leaders Fellowship [Grant MR/T040866/1], a Science and Technology Facilities Funding Council Small Award [Grant ST/T000082/1], and the Leverhulme Trust through a research project grant [RPG-2020-82].
Harry Baskett acknowledges support from a Science and Technology Facilities Council Studentship [ST/Y509383/1].
Duncan A. Christie is supported by the Max Planck Society.
Mercedes López-Morales is supported by individual research time under NASA contracts NAS5-26555 and NAS5-03127 to the Associated Universities for Research in Astronomy for the operation of the Hubble Space Telescope and the James Webb Telescope Science Operations Centers at STScI.
This work is based in part on observations made with the NASA/ESA/CSA JWST and HST. The data were obtained from the Mikulski Archive for Space Telescopes at the Space Telescope Science Institute, which is operated by the Association of Universities for Research in Astronomy, Inc., under NASA contract NAS 5-03127 for JWST.
The specific JWST NIRISS observation analyzed here can be accessed via \dataset[DOI:10.17909/bkfz-6x88]{https://doi.org/10.17909/bkfz-6x88}, and the HST WFC3/UVIS G280 observation can be accessed via \dataset[DOI:10.17909/0tkq-yk07]{https://doi.org/10.17909/0tkq-yk07}.
This work has made use of data from the European Space Agency (ESA) mission Gaia (\url{https://www.cosmos.esa.int/gaia}), processed by the Gaia Data Processing and Analysis Consortium (DPAC, \url{https://www.cosmos.esa.int/web/gaia/dpac/consortium}).  Funding for the DPAC has been provided by national institutions, in particular the institutions participating in the Gaia Multilateral Agreement.
The GCM results are produced using Met Office Software and the Monsoon3 system, a collaborative facility supplied under the Joint Weather and Climate Research Programme, a strategic partnership between the Met Office and the Natural Environment Research Council in the UK.
This publication makes use of data products from the Two Micron All Sky Survey, which is a joint project of the University of Massachusetts and the Infrared
Processing and Analysis Center/California Institute of Technology, funded by the National Aeronautics and Space Administration and the National Science Foundation.  This publication makes use of data products from the Wide-field Infrared Survey Explorer, which is a joint project of the University of California, Los Angeles, and the Jet Propulsion Laboratory/California Institute of Technology, funded by the National Aeronautics and Space Administration.  
This research has made use of the SIMBAD database, operated at CDS, Strasbourg, France \citep{wen00}.
This research has made use of the VizieR catalog access tool, CDS, Strasbourg, France.  The original description of the VizieR service was published in \citet{och00}. 
This research has made use of the NASA Exoplanet Archive, which is operated by the California Institute of Technology, under contract with the National Aeronautics and Space Administration under the Exoplanet Exploration Program. 
This research has made use of NASA's Astrophysics Data System Bibliographic Services.

\facilities{ADS, CDS, CTIO:2MASS, Exoplanet Archive, FLWO:2MASS, Gaia, GALEX, HST(WFC3), JWST(NIRISS, NIRSpec), MAST, Simbad, WISE}

\software{ \vspace{0.1cm}
\\ \texttt{AEOLUS} \citep{Aeolus_2024},
\\ \texttt{astropy} \citep{AstropyI, AstropyII, AstropyIII},
\\ \texttt{batman} \citep{Kreidberg2015},
\\ \texttt{catwoman} \citep{Jones22, Espinoza21},
\\ \texttt{dustmaps} \citep{gre18},
\\ \texttt{emcee} \citep{Foremak-Mackey2013},
\\ \eureka \citep{Eureka!},
\\ \texttt{ExoTiC-LD} \citep{grant2024exoticLDJoss},
\\ \firefly \citep{Rus22, Rus23},
\\ \texttt{gaiadr3\_zeropoint} \citep{lin21a},
\\ IPython \citep{ipython},
\\ \texttt{IRIS} \citep{Iris_2026},
\\ \texttt{isochrones} \citep{mor15},
\\ \texttt{lacosmic} \citep{lacosmic},
\\ \texttt{lmfit} \citep{lmfit},
\\ \texttt{matplotlib} \citep{hunter2007matplotlib},
\\ \texttt{numpy} \citep{harris2020array},
\\ \picaso \citep{Batalha19, Mukherjee23}
\\ \poseidon \citep{MacDonald2017, MacDonald2023},
\\ \texttt{PyMultiNest} \citep{Feroz2009,Buchner2014},
\\ \texttt{scipy} \citep{jones2001scipy, 2020SciPy-NMeth}
}

\bibliography{article_bibilography}{}
\bibliographystyle{aasjournal}


\appendix

\section{Demonstration of Charge Migration in NIRSpec PRISM}\label{app:nirspec}

\begin{figure*}
    \centering
    \includegraphics[width=\linewidth]{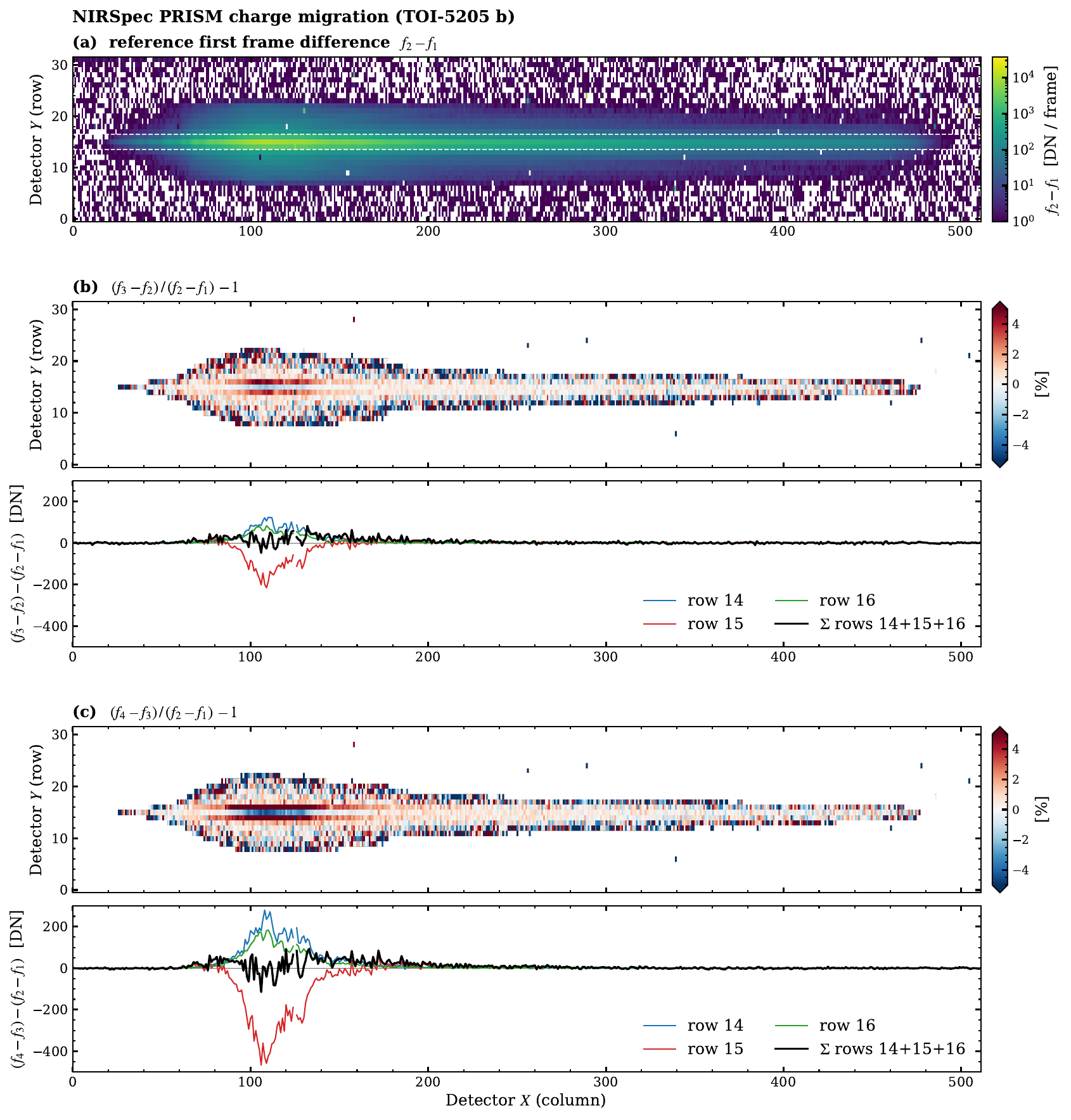}
    \caption{Demonstration of charge migration in the NIRSpec PRISM detector during the observation of TOI-5205\,b \citep[][available at \url{https://doi.org/10.17909/29st-dz13}]{Canas26}.
    In panel \textbf{(a)}, we show the difference between the first two frames as a point of reference, and plot as dotted horizontal white lines the NIRSpec PRISM trace.
    In panels \textbf{(b)} and \textbf{(c)} we show the relative nonlinearity following the same calculation as in Figure \ref{fig:chargemigration_avg}, but with the third frame \textbf{(b)} and the fourth frame \textbf{(c)}.
    The lower panels of \textbf{(b)} and \textbf{(c)} show the relative nonlinearity of the rows near the middle of the trace, with the sum of the three as the black line.
    The same structure of low and high relative nonlinearity that follows the PSF peak gradient seen in NIRISS is also present in NIRSpec data, motivating the use of our late-ramp-fit methodology for the other JWST instruments.
    }
    \label{fig:chargemigration_nirspec}
\end{figure*}
 
In addition to NIRISS, we expect that the other JWST instruments' detectors should exhibit charge migration as we have shown in Section \ref{sec:chargemigration}.
As an example, we show in Figure \ref{fig:chargemigration_nirspec} a relative nonlinearity figure similar to Figure \ref{fig:chargemigration_avg} for the NIRSpec PRISM observations of TOI-5205.
Like NIRISS, these observations exhibit alternating rows of positive and negative relative nonlinearity indicative of charge migration, which largely cancels when summed together.
This suggests that our late-ramp-fit methodology will be broadly applicable to JWST time series observations with counts close to the detector limits.

\section{Atmosphere Model Comparison} \label{app:opacities}

\begin{deluxetable*}{lccc} \label{tab:opacities}
    \centering
    \tablewidth{0pt}
    \tablecaption{Model Opacity/Line List Sources and Retrieval Configuration}
    \tablehead{
    Molecule & \colhead{\poseidon} & \colhead{\pRT} & \colhead{\picaso}
    }
    \startdata
    H$_2$O & \citet{polyansky18} & \citet{polyansky18} & \citet{polyansky18} \\
    CH$_4$ & \citet{Yurchenko2024} &--- &--- \\
    CO & \citet{li15} & \citet{li15} & \citet{li15},\\
     & & & \citet{Rothman:2010},  \\
     & & & \citet{HITRAN2016},\\
    CO$_2$ &  \citet{yurchenko20} &  \citet{yurchenko20} & \citet{Huang14} \\
    HCN & \citet{barber14} & \citet{harris06} & --- \\
    H$_2$S & \citet{azzam16exomol} & \citet{azzam16exomol} & \citet{Azzam15} \\
    SO$_2$ & \citet{Underwood2016} &--- & ---\\
    OCS & \citet{Owens2024} &--- & \citet{Rothman13} \& \\
     & & & \citet{Wilzewski16}\\
    TiO & \citet{McKemmish19} &--- & \citet{McKemmish19} \&  \\
     & & & \citet{GharibNezhad2021}\\
    VO & \citet{McKemmish16} &--- & \citet{McKemmish16} \& \\
     & & & \citet{GharibNezhad2021}\\
    SiO & \citet{Yurchenko:2022} &--- &--- \\
    AlO &  \citet{Patrascu15}&--- &--- \\
    SH & \citet{Gorman19} &--- & ---\\
    FeH & \citet{Lodi15} &--- & ---\\
    NH$_3$ & ---& \citet{Gordon2022} & --- \\
    K &--- & \citet{Molliere19} & ---\\
    Na &--- & \citet{Allard2019} & ---\\
    H$^-$ & \citet{John88} & --- & \citet{John88} \\
    \enddata
\end{deluxetable*}

\begin{deluxetable*}{lcc}
\centering
\tablewidth{0pt}
\tablecaption{Atmospheric Retrieval Configuration Comparison}
\tablehead{
Model Setting & \colhead{\poseidon} & \colhead{\pRT}
}
\startdata
    \hline
    \multicolumn{3}{c}{\textbf{Atmospheric Model}}\\
    \hline
    Pressure Grid & $10^{-7}$--$10^2$\,bar & $10^{-6}$--$10^3$\,bar  \\
    Number of Layers & 150 & 80 \\
    Molecules Included & H$_2$O, CH$_4$, CO, CO$_2$, HCN,  & H$_2$O, CO, CO$_2$, NH$_3$, \\
    & SO$_2$, H$_2$S, OCS, TiO, VO, & H$_2$S, HCN, K, Na \\
    & SiO, AlO, FeH, SH, H$^-$ &  \\
    He/H$_2$ Ratio & 0.17 & 0.17 \\
    P-T Profile Treatment & \citet{Madhusudhan2009} & \citet{Guillot10}\\
    Cloud Treatment & Mie Scattering & Opaque Cloud \\
    \hspace{1pt} (citation) & \citep{Mullens24} & --- \\
    Hydrostatic Boundary Condition & $\log_{\rm{10}} P_{\rm{ref}}$ & $\log_{\rm{10}} P_{\rm{ref}}$ \\
    Stellar Contamination Model & \citet{Rathcke2021} One-spot & --- \\
    Data Used & HST WFC3/UVIS, JWST NIRISS, & JWST NIRISS, JWST \\
     & JWST NIRSpec/G395H & NIRSpec/G395H\\
    \hline
    \multicolumn{3}{c}{\textbf{Spectral Model}}\\
    \hline
    Model Wavelength Grid & 0.2--5.3\,$\mu\text{m}$ & 0.85-5.2\,$\mu\text{m}$ \\
    Native Opacity Resolution & 0.01\,cm$^{-1}$ & $R = $ 1,000 \\
    Opacity Sampling Resolution & $R = $ 20,000 & --  \\
    \hline
    \multicolumn{3}{c}{\textbf{Retrieval Priors and Settings}}\\
    \hline
    Planetary Surface Gravity ($\log_{\rm{10}} g_{\rm{p}}$) &  3.09 (fixed) & 3.129 (fixed) \\
    Reference Pressure ($\log_{\rm{10}} (P_{\rm{ref}}$ / bar)) & 10$^{-3}$ (fixed) & 10$^{-1}$ (fixed) \\
    Reference Radius ($\text{R}_{\mathrm{p, \, ref}}$) & $\mathcal{N}(1.56, 0.31^2)\,\text{R}_{\text{Jup}}$ & $\mathcal{U}(1.25, 2.00)\,\text{R}_{\text{Jup}}$ \\
    \hline
    Atmospheric Temperature ($T_{\mathrm{ref}}$) & $\mathcal{U}(750, 2500)$\,K & $\mathcal{U}(1000, 4000)$\,K \\
    $T(P)$ Profile Curvature 1 ($\alpha_{1}$) & $\mathcal{U}(0.3, 2.00)$\,K$^{-\frac{1}{2}}$ & ---  \\
    $T(P)$ Profile Curvature 2 ($\alpha_{2}$) & $\mathcal{U}(0.3, 2.00)$\,K$^{-\frac{1}{2}}$ & ---  \\
    $T(P)$ Profile Region 1 ($\log_{10} (P_{1}$ / bar)) & $\mathcal{U}(-6, 0)$ & --- \\
    $T(P)$ Profile Region 2 ($\log_{10} (P_{2}$ / bar)) & $\mathcal{U}(-6, 0)$ & --- \\
    $T(P)$ Profile Region 3 ($\log_{10} (P_{3}$ / bar)) & $\mathcal{U}(-2, 2)$ & --- \\
    Infrared Opacity ($\log_{10} (\kappa_{\text{IR}}$ / cm$^2$ g$^{-1})$) & --- & $\mathcal{U}(-4, 2)$ \\
    Optical/Infrared Opacity Ratio ($\log_{10} (\gamma)$) & --- & $\mathcal{U}(-3, 3)$ \\
    Intrinsic Temperature ($T_{\mathrm{int}}$) & --- & 500 K (Fixed) \\
    \hline
    Cloud Top Pressure ($\log_{\rm{10}} (P_{\rm{cloud}}$ / bar)) & $\mathcal{U}(-6, 1)$ & $\mathcal{U}(-6, 2)$ \\
    Aerosol Volume Mixing Ratio ($\log_{\rm{10}} \text{Al}_2\text{O}_3$) & $\mathcal{U}(-30, -1)$ & --- \\
    Aerosol Particle Size ($\log_{\rm{10}} (r_{\text{Al}_2\text{O}_3}$/$\mu$m)) & $\mathcal{U}(-3, -1)$ & --- \\
    Slab Pressure Extent ($\Delta\log_{10} P_{\text{slab}}$) & $\mathcal{U}(0, 7)$ & --- \\
    \hline
    Stellar Photosphere Temperature $T_{\rm{phot}}$ & $\mathcal{N}(6632, 60^{2})$\,K & --- \\
    Stellar Heterogeneity Temperature $T_{\rm{het}}$ & $\mathcal{U}(4738, 8122)$\,K & ---  \\
    Stellar Heterogeneity Fraction $f_{\rm{het}}$ & $\mathcal{U}(0, 0.5)$ & ---  \\
    \hline
    Free Chem Molecule Volume Mixing Ratio ($\log_{\rm{10}} X_i$) & $\mathcal{U}(-12, -0.3)$ & --- \\
    Equilibrium Chem Metallicity ($\log_{\rm{10}} \text{[M/H]}$) & $\mathcal{U}(0.5, 4.0)$ & $\mathcal{U}(-1.0, 3.0)$ \\
    Equilibrium Chem C/O & $\mathcal{U}(0.2, 2.0)$ & $\mathcal{U}(0.1, 1.6)$ \\
    \hline
    Data Offset 1 ($\delta_{\mathrm{rel, \, 1}}$) & $\mathcal{U}(-750, 750)$\,ppm & $\mathcal{U}(-1000, 1000)$\,ppm  \\
    Data Offset 2 ($\delta_{\mathrm{rel, \, 2}}$) & $\mathcal{U}(-750, 750)$\,ppm & $\mathcal{U}(-1000, 1000)$\,ppm  \\
    Data Offset 3 ($\delta_{\mathrm{rel, \, 3}}$) & $\mathcal{U}(-750, 750)$\,ppm & ---  \\
    \hline
    \texttt{MultiNest} Live Points & 1000 & 1000  \\
    Retrieval Code Availability & \href{https://github.com/MartianColonist/POSEIDON}{\poseidon GitHub} & \href{https://gitlab.com/mauricemolli/petitRADTRANS}{\pRT GitLab} \\
    \hline
\enddata
\tablecomments{\pRT uses the planetary surface gravity from \citet{Ahrer25} to facilitate a direct comparison with that analysis.
Gaussian priors are summarized as $\mathcal{N}(\mu, \sigma^2)$, where $\mu$ and $\sigma$ are the mean and standard deviation, respectively. 
$T_{\mathrm{ref}}$ refers to the top-of-atmosphere temperature for \poseidon, while for \pRT it represents the equilibrium temperature.
As \pRT lacks a stellar photosphere inhomogeneity treatment, our retrievals with it only consider the JWST NIRISS and JWST NIRSpec data.
}
\label{tab:retrieval_configurations}
\end{deluxetable*}

In Table \ref{tab:opacities} we enumerate the opacity sources used in the Bayesian inference of KELT-7\,b's atmosphere in Section \ref{sec:retrievals}. We then describe the retrieval configurations in Table \ref{tab:retrieval_configurations}.
The majority of the overlapping molecules have the same sources across the three codes.

\section{\pRT Line List Comparison} \label{app:pRTlinelists}
\begin{figure*}
    \centering
    \includegraphics[width=0.7\linewidth]{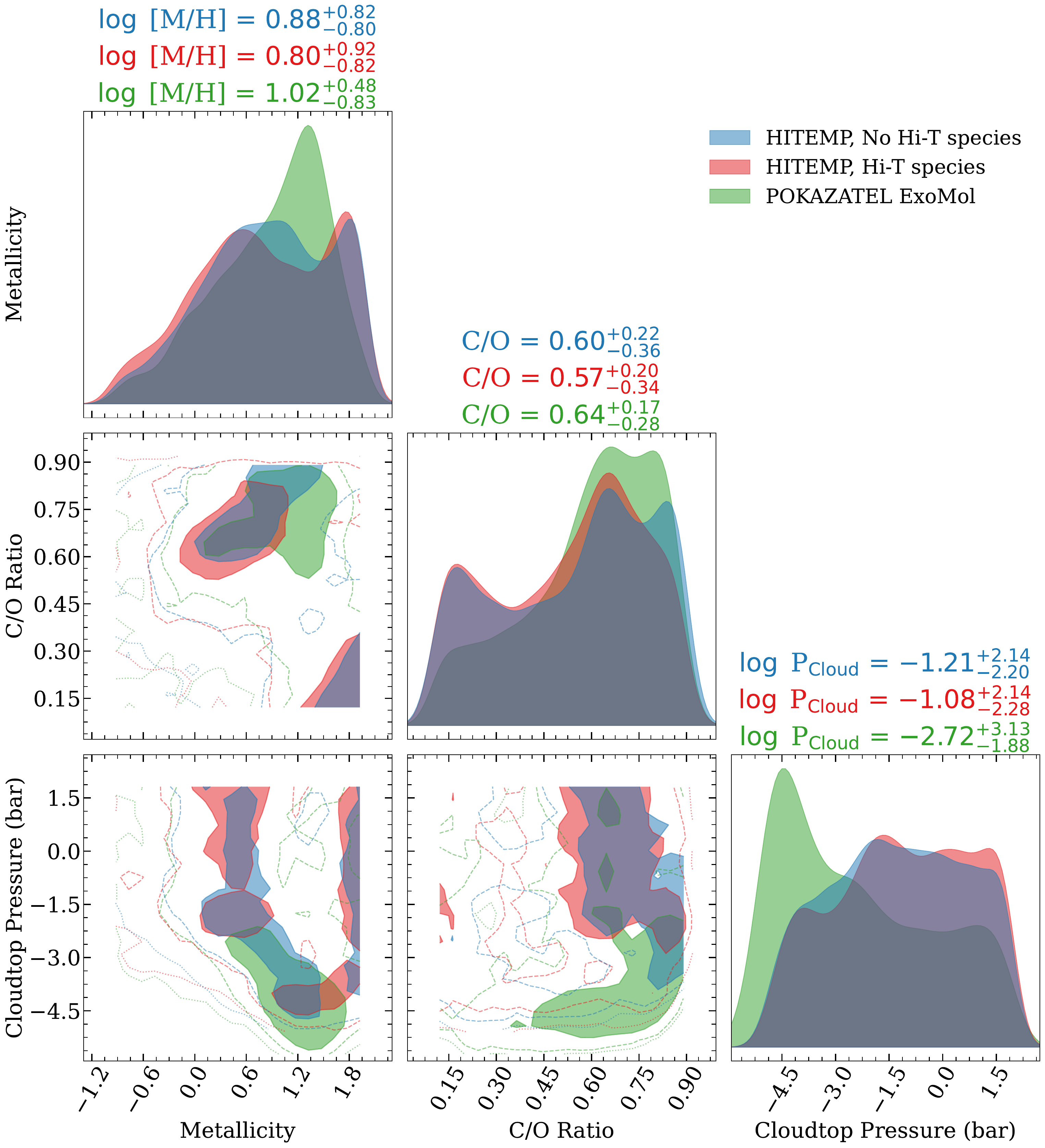}
    \caption{Comparison of posterior distributions of our \pRT retrievals with differing line lists on the JWST NIRSpec/G395H observations presented in \citet{Ahrer25}. When we use the ExoMol POKAZATEL line list for H$_2$O \citep{polyansky18}, we can recreate the results of \citet{Ahrer25}'s \pRT retrievals. The differing results from our retrievals on the combination of NIRISS and NIRSpec therefore reflect the additional information from NIRISS rather than a difference in model inputs.}
    \label{fig:nirspec_list_compare}
\end{figure*}
While attempting to recreate the results of \citet{Ahrer25}'s \pRT retrieval analysis, we found that the atmospheric constraints were sensitive to the choice of line lists. 
We originally used HITEMP as our source for H$_2$O opacity data \citep{Rothman:2010} and included an expanded list of high-temperature species, including SiO \citep{Yurchenko:2022}, TiO \citep{McKemmish19}, and VO \citep{McKemmish16}. 
Our [M/H], C/O, and $P_{\rm cloud}$ measurement were far less constraining than in \cite{Ahrer25}, with long tails towards both high and low metallicity and C/O. 
Removing the high-temperature species only had a small effect on the posterior distributions. 
We then tested replacing the HITEMP H$_2$O line list with the ExoMol POKAZATEL list \citep{polyansky18}, resulting in posteriors that matched \cite{Ahrer25} to $<0.2\sigma$. 
This demonstrates that we can recreate the results of \citet{Ahrer25} with an identical retrieval setup; therefore, our results when including NIRISS/SOSS data reflect the additional information from those observations rather than a difference in retrieval setup.

\section{Equilibrium Retrieval/Bayesian Analysis Corner Plots} \label{app:eqcorner}
\begin{figure*}
    \centering
    \includegraphics[width=\linewidth]{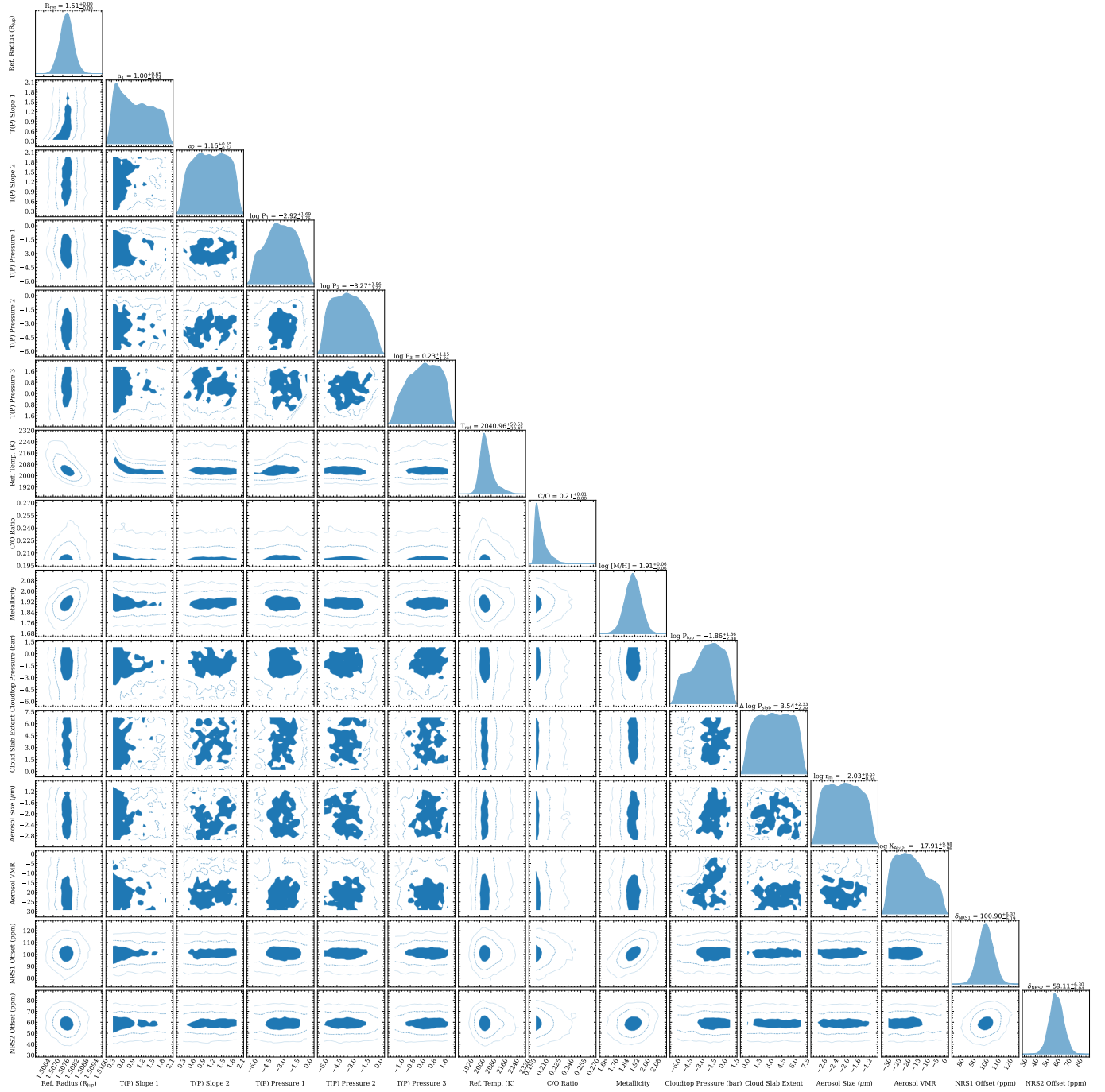}
    \caption{Corner plot showing the posterior of our \poseidon equilibrium chemistry retrieval. The solid regions in each covariance plot represent the 1-$\sigma$ region of the posterior; likewise, the solid and dashed lines represent the 2-$\sigma$ and 3-$\sigma$ regions. More details about the priors are available in Table \ref{tab:retrieval_configurations}.}
    \label{fig:poseidon_corner}
\end{figure*}

\begin{figure*}
    \centering
    \includegraphics[width=\linewidth]{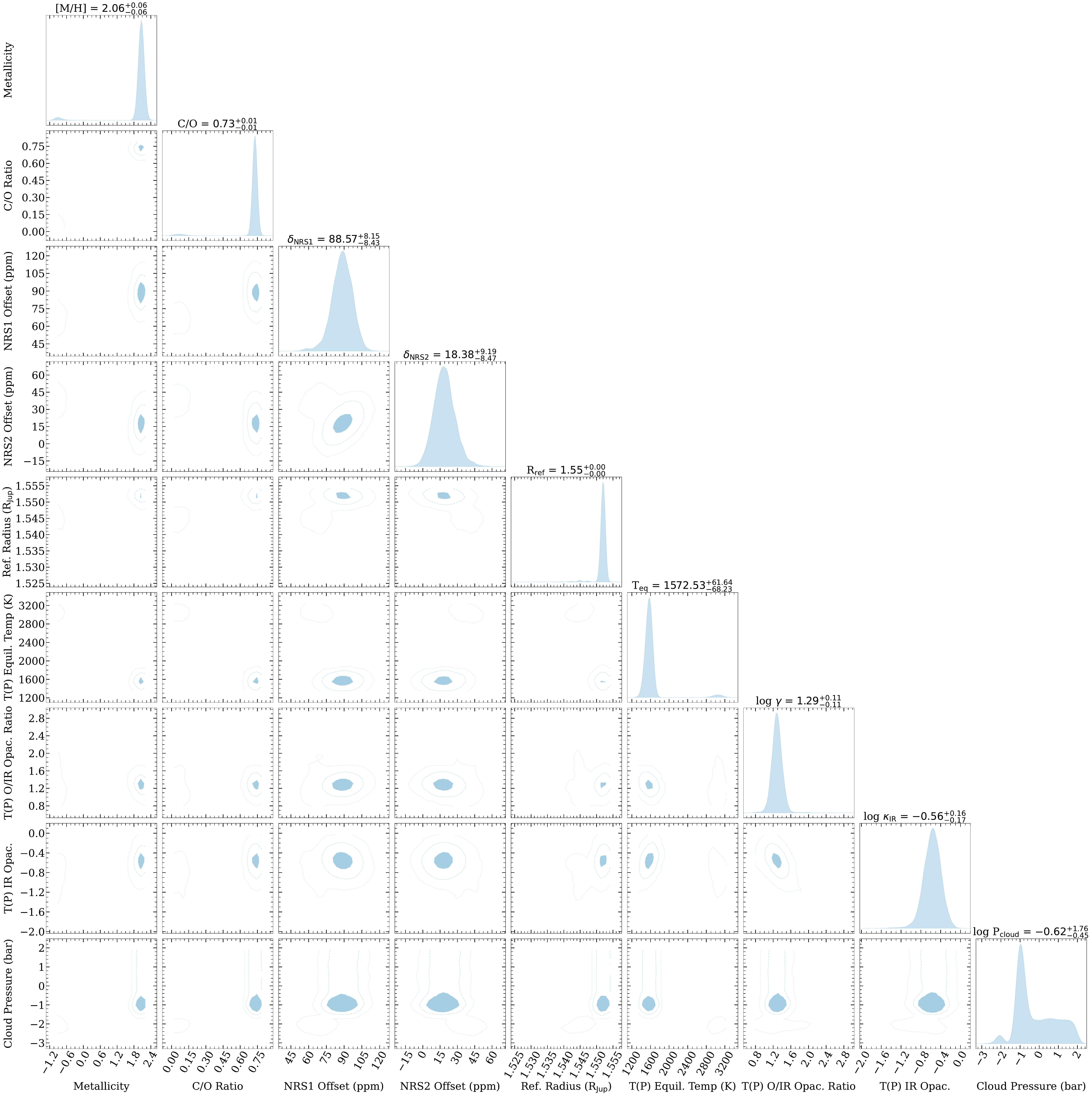}
    \caption{Corner plot showing the posterior of our \pRT NIRISS \& NIRSpec equilibrium chemistry retrieval. More details about the priors are available in Table \ref{tab:retrieval_configurations}.}
    \label{fig:prt_corner}
\end{figure*}

\begin{figure*}
    \centering
    \includegraphics[width=\linewidth]{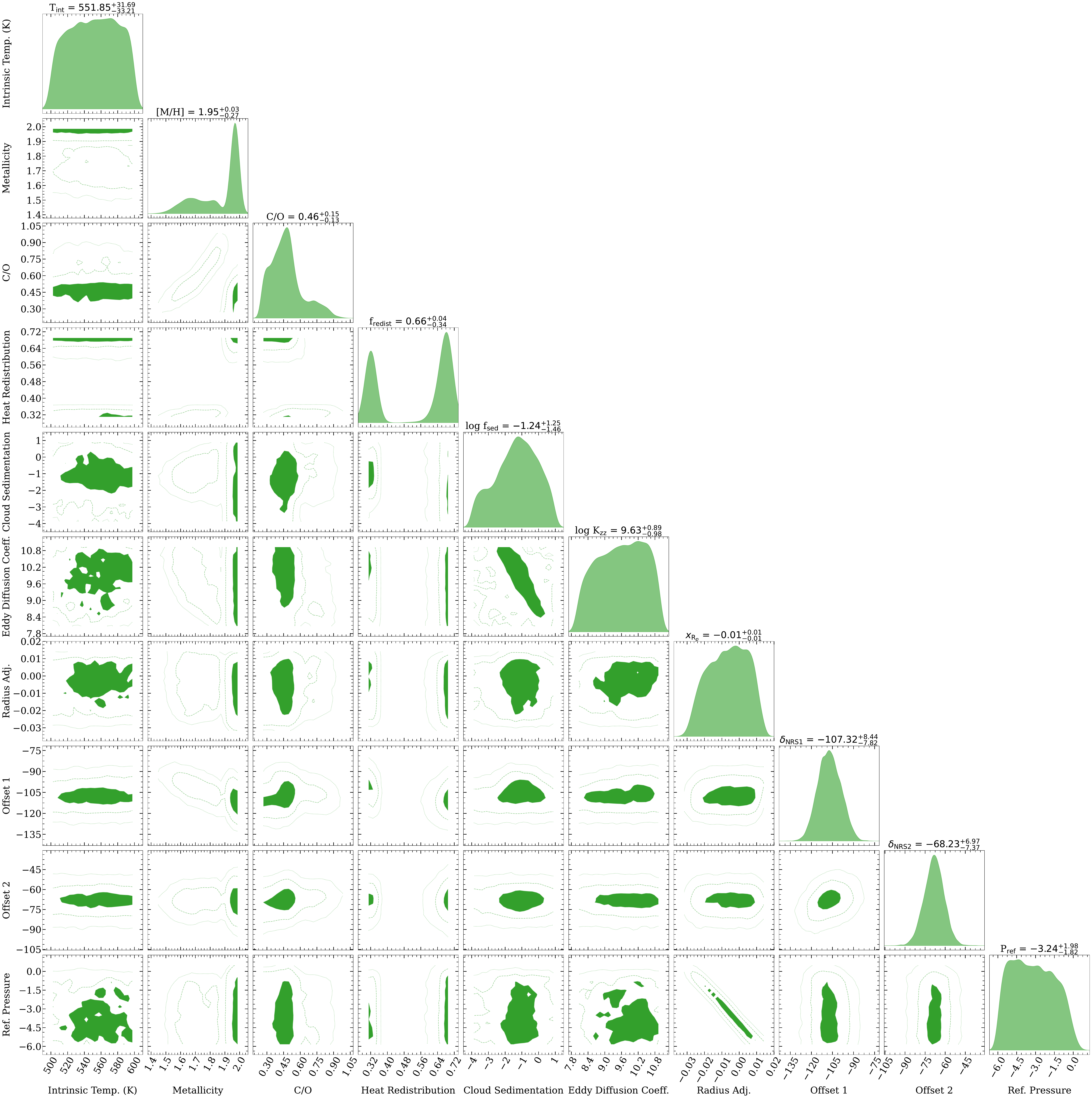}
    \caption{Corner plot showing the posterior of our \picaso Bayesian grid fitting analysis.}
    \label{fig:picaso_corner}
\end{figure*}

Here we show for documentation purposes the corner plots for our equilibrium retrievals/Bayesian grid fitting analysis: Figure \ref{fig:poseidon_corner} for \poseidon, Figure \ref{fig:prt_corner} for \pRT, and Figure \ref{fig:picaso_corner} for \picaso.

\end{document}